\documentclass[]{aa}
\usepackage[dvipsnames]{xcolor}
\usepackage{soul}
\usepackage{ulem}
\usepackage{graphicx}
\usepackage[varg]{txfonts}
\usepackage[]{hyperref}
\usepackage[flushleft]{threeparttable}
\usepackage{natbib}
\usepackage{verbatim}
\usepackage{float}
\bibpunct{(}{)}{;}{a}{}{,}

\def\vw{\upsilon_{\rm wind}}
\def\vorb{\upsilon_{\rm orb}}
\def\ro{R_{\rm O}}
\def\rd{R_{\rm disk}}
\def\risco{R_{\rm ISCO}}
\def\mbh{M_{\rm BH}}
\def\mo{M_{\rm O}}
\def\ms{M_\odot}
\def\rdrisco{ the ratio of the circularisation radius to the radius of the innermost stable circular orbit}
\def\simgr{\mathrel{\hbox{\rlap{\hbox{\lower4pt\hbox{$\sim$}}}\hbox{$>$}}}}
\begin{document}

\title{X-ray emission from BH+O star binaries expected to descend from the observed galactic WR+O binaries}


\author{K. Sen\inst{1,2}
\thanks{email: ksen@astro.uni-bonn.de}
\and X.-T. Xu\inst{1,2}
\thanks{The first two authors have contributed equally to this work}
\and N. Langer\inst{1,2}
\and I. El Mellah\inst{3}
\and C. Sch\"{u}rmann\inst{1,2}
\and M. Quast\inst{1}
}

\institute{Argelander-Institut fur Astronomie, Universitat 
Bonn, Auf dem Hugel 71, 53121 Bonn, Germany
\and Max-Planck-Institut fur Radioastronomie, Auf dem Hugel 
69, 53121 Bonn, Germany
\and Univ. Grenoble Alpes, CNRS, IPAG, 414 Rue de la Piscine, 38400 Saint-Martin-d'Hères, France
}

\date{Received \today / Accepted ...}

\abstract {In the Milky Way, $\sim$18 Wolf-Rayet+O star (WR+O) binaries 
are known with estimates of their stellar and orbital parameters.
Whereas black hole+O star (BH+O) binaries are thought to evolve from WR+O 
binaries, only one such system is known in the Milky Way. To resolve this 
disparity, it was suggested recently that upon core collapse, the WR 
stars receive large kicks such that most of the binaries are disrupted.}
{We reassess this issue, with particular emphasis on the uncertainty 
in predicting the X-ray emission from wind-accreting BHs in BH+O 
binaries, which is key to identifying such systems.}
{BH+O systems are thought to be X-ray bright only when an accretion 
disk forms around the BHs. We follow the methodology of previous work 
and apply an improved analytic criterion for the formation of an
accretion disk around wind accreting BHs. We then use stellar 
evolutionary models to predict the properties of the BH+O binaries 
which are expected to descend from the observed WR+O binaries if the 
WR stars would form BHs without a natal kick.}
{We find that disk formation depends sensitively on the O stars' 
wind velocity, the amount of specific angular momentum carried by 
the wind, the efficiency of angular momentum accretion by the BH, 
and the spin of the BH. We show that whereas the assumption of a 
low wind velocity may lead to the prediction that most of the BH+O star 
binaries will have an extended X-ray bright period, this is not the 
case when typical wind velocities of O stars are considered. We find 
that a high spin of the BH can boost the duration of the X-ray 
active phase as well as the X-ray brightness during this phase. 
This produces a strong bias for detecting high mass BH binaries 
in X-rays with high BH spin parameters.}
{We find that large BH formation kicks are not required to understand
the sparsity of X-ray bright BH+O stars in the Milky Way. Probing for 
a population of X-ray silent BH+O systems with alternative methods can 
likely inform us about BH kicks and the necessary conditions for high 
energy emission from high mass BH binaries.
}

\keywords{Stars: massive -- Stars: evolution -- Stars: black holes -- X-rays: binaries -- (Stars) binaries: close}

\titlerunning{Detectability of BH+O binaries in the Milky Way}
\authorrunning{K. Sen et al.}

\maketitle

\section{Introduction}

The detection of gravitational waves by LIGO/VIRGO in the last decade 
has opened a new window to look at our Universe. Since the first 
observation by LIGO in 2015 \citep{abbott2016,abbott2019}, most of these, now 
routine, events are associated with merging stellar mass black holes \citep{abbott2019}. 
Thereby, the interest in the study of black holes has been revitalised 
\citep{selma2016,Marchant2016,belczynski2020,woosley2020,Buisson2020}. 
But the evolution of massive star binaries towards binary compact object mergers 
is still riddled with uncertainties \citep{langer2012,crowther2019}.

Apart from gravitational wave signals from compact object mergers and 
direct imaging of the supermassive BH shadows \citep{eht2019}, BHs can be 
detected via microlensing \citep{minniti2015,masuda2019,wyrzykowski2020}, 
tidal disruption events \citep{perets2016,kremer2019}, and X-ray emission 
due to accretion onto the BH. In the latter case, the source of material 
can be dense interstellar medium \citep{fujita1998,tsuna2018,Scarcella20}, 
or an orbiting stellar companion \citep{orosz2011}.

A large number of binary population synthesis studies have been 
undertaken to predict the event rate of merging compact objects 
\citep{mennekens2014,belczynski2014,stevenson2015,demink2015,kruckow2018}. 
One of the major uncertainties in population synthesis studies 
\citep[for a discussion, see][]{shaughnessy2008} 
is whether the formation of a BH is preceded by a SN 
explosion and if so, whether the BH receives a natal kick high 
enough to disrupt the binary in which the BH formed 
\citep{mandel2020a,mandel2020b,woosley2020}. As expected, 
the presence or absence of a substantial kick during BH 
formation significantly affects the BH-BH merger rates 
calculated by population synthesis calculations 
\citep{mennekens2014,belczynski2016}. 

Direct evidence towards high or low BH kicks is inconclusive 
\citep{ozel2010,farr2011,belczynski2012}. On one hand, in Galactic low mass 
X-ray binaries containing a BH, BHs were found to have formed with low or 
modest kick velocities \citep{brandt1995,willems2005,fragos2009,wong2012}. 
\citet{belczynski2016} (table 7, and references therein) have 
given empirical evidence for low BH natal kicks. On the other hand, 
\citet{repetto2012,repetto2017} found that their binary models can 
adequately explain the observed population of low mass BH binaries 
above the Galactic plane when high BH kick velocities, similar to 
the ones assumed for the formation of neutron stars \citep{hobbs2005}, are 
adopted during BH formation. Moreover, some works have suggested a BH 
mass dependent natal kick distribution \citep{mirabel2003,dhawan2007}, 
with more massive BHs receiving lower kicks.

Several teams have studied whether 
very massive stars can explode at the end of their lifetime 
\citep{oconner2011,ugliano2012}. \citet{sukhbold2018} and 
\citet{woosley2019} predict that most of the hydrogen free 
Helium stars having masses between 7-30 M$_{\odot}$, which 
will also manifest as Wolf-Rayet stars during helium burning, 
do not explode with an associated supernova but instead 
implode into BHs. \citet{mirabel2003} show evidence that WR 
stars might become BHs with little or no kick. 

\citet{langer2020} predicted to find $\sim$3 out of every 100 massive 
binary stars to host a BH. The average lifetime of the WR+O phase 
($\sim$0.4 Myrs, given by the lifetime of the WR phase) is much smaller 
than the lifetime of the BH+O phase (which is given by the remaining 
main sequence lifetime of the O star). Hence, if the transition from 
the WR+O stage to the BH+O stage happens without disruption of the binary, 
we expect the Milky Way to host more binaries containing BHs than WR 
stars. But the observed number of WR+O star binaries are much larger 
than BH+O star binaries.

\citet{vanbeveren2020} (hereafter V20) assessed this problem with 
the following two assumptions: (i) WR stars collapse to form BHs 
with no natal kick, (ii) a BH+O binary is detectable if the BH has 
an accretion disk and the X-ray flux emitted from the accretion 
disk is above the detection threshold of current X-ray telescopes. 
They predicted to find over 200 wind-fed BH high mass X-ray Binaries 
(HMXBs) in the Milky Way. There is only one observed in the Milky Way 
\citep[Cygnus X-1, see e.g.][]{Hirsch19}.

The large discrepancy between the predicted and observed number of 
wind-fed BH HMXBs led V20 to conclude that most of the WR stars must 
explode in a supernova to form neutron stars with an associated large 
natal kick that disrupts the binaries, or BH formation itself is 
associated with a high kick velocity that disrupts most of the progenitor 
WR+O binaries at the time of BH formation. This conclusion would greatly 
affect the merger rates of BH-BH and BH-NS mergers as many population 
synthesis results assume low kick velocities for BH formation. 

In this work, we follow \citet{shapiro1976} to formulate a condition 
for the formation of accretion disks and detectability of a BH+O 
system as a wind-fed BH HMXB. We investigate the effect of the stellar 
wind velocity, efficiency of angular momentum accretion from the stellar 
wind and the spin of the BH, on our prediction of the number of wind-fed 
BH HMXBs. We also revisit the assumptions and definitions of stellar 
parameters used to derive the accretion disk formation criterion in the 
work of V20. 

In Sect.\,\ref{method}, we outline the definitions and assumptions used 
to derive our accretion disk formation criterion. We then predict the 
population of BH+O binaries and study the effect of uncertain parameters 
on our predictions in Sect.\,\ref{results}. We compare the assumptions 
and results in our work with the literature in Sect.\,\ref{earlier_work}. 
In Sect.\,\ref{dnc}, we critically discuss the implications of the uncertainties 
that are present in the calculation of the X-ray active lifetime of BH+O 
binaries and outline our main conclusions from this work in Sect.\,\ref{conclusion}.

\begin{table}[H]
\centering
\caption{Stellar parameters of the anticipated BH+O binaries obtained by V20 at BH formation, in order of increasing orbital period. The BH is assumed to be formed at the end of core helium depletion of the WR star in the progenitor WR+O binaries.
}
\begin{tabular}{l r r r r r}
\hline
\hline
Progenitor & distance    & O star        & BH            & Orbital & $L/L_{\rm Edd}$    \\    
System     &             & mass          & mass          & period   & of O star      \\   
           & (kpc)       & ($M_{\odot}$) & ($M_{\odot}$) & (days)         \\   
\hline
WR 155 & 2.99 & 30 & 12  &   2.6 & 0.161   \\
WR 151 & 5.38 & 28 & 10  &   3.4 & 0.076   \\
WR 139 & 1.31 & 28 & 6   &   5.0 & 0.101  \\
WR 31  & 6.11 & 24 & 7   &   6.1 & 0.140   \\
WR 42  & 2.44 & 27 & 14  &   8.7 & 0.156   \\
WR 47  & 3.49 & 47 & 20  &  10.5 & 0.317   \\
WR 79  & 1.37 & 24 & 7   &  10.7 &  0.076  \\
WR 127 & 3.09 & 20 & 6   &  11.8 & 0.076   \\
WR 21  & 3.99 & 37 & 10  &  11.8& 0.341    \\
WR 9   & 4.57 & 32 & 8   &  15.0& 0.299    \\
WR 97  & 2.15 & 30 & 9   &  18.3 & 0.304   \\
WR 30  & 5.09 & 34 & 14  &  20.4 & 0.303   \\
WR 113 & 1.80 & 22 & 8   &  35.9 &  0.054  \\
WR 141 & 1.92 & 26 & 18  &  43.1 & 0.076   \\
WR 35a & 5.84 & 19 & 10  &  68.2 & 0.054   \\
WR 11  & 0.34 & 31 & 8   &  86.8& 0.107    \\
WR 133 & 1.85 & 34 & 9   & 158.0 & 0.107   \\
\hline
\end{tabular}
\label{comparison}
\end{table}

\section{Method}
\label{method}
\subsection{Sample selection}
\label{sample}

In the Milky Way, there are about $\sim$53 observed WR+O type 
binaries\footnote{http://pacrowther.staff.shef.ac.uk/WRcat/index.php}
\citep{hucht2001,hucht2006,crowther2015,rosslowe2015}. Of 
them, 38 are designated as double-lined spectroscopic binaries (SB2). V20 
consider a sub-population of 17 SB2 binaries that have estimates 
of the masses of both components and orbital period of the binary. 
The present masses of both components and the orbital period 
of the selected sample of 17 binaries can be found in table 1 of V20. 
We find one more SB2 system, WR 22, that has estimates of 
its component masses and orbital period \citep{Schweickhardt1999}. 
This system has an orbital period around $\sim$80 days. In this 
work, we further look at the distance of the systems from Earth using the 
catalogue of galactic WR stars \citep{rosslowe2015} (Table \ref{comparison}). 
To be consistent with the analysis of V20, we choose to analyse the 
sub-sample of the 17 WR+O binaries. We also explain later that the 
addition of WR 22 to the sample of 17 SB2 binaries reinforces the 
conclusions we derive from our work. 

The orbital period distribution of WR+O binaries in the Large 
Magellanic Cloud is expected to peak at $\sim$100 days \citep{langer2020}, 
which can be expected to be similar in the Milky Way. Observationally, 
short period WR+O star binaries are much easier to detect than long period 
ones. This implies that the sub-sample of $\sim$17 mostly short period 
WR+O binaries considered in this work may indeed account for nearly all short 
period WR+O binaries expected for the $\sim$53 WR+O binaries observed in the Milky 
Way. We will see below that only short period WR+O binaries can manifest as 
X-ray bright BH+O systems. In this sense, the sub-sample of 17 WR+O binaries 
can be used as a suitable proxy to analyse the detectability of anticipated 
BH+O binaries in the Milky Way.

\subsection{Binary evolution}

We describe the further modelling of the chosen WR+O binaries in the 
following subsections.

\subsubsection{WR+O binary evolution up to BH formation}

We adopt the stellar and orbital parameters of the anticipated BH+O 
binaries derived by V20 at the time of BH formation (Table \ref{comparison}). 
We describe the modelling of the evolution of the WR+O star binaries 
performed by V20 up to the point of BH formation briefly in the 
following paragraph. 

The orbital periods of the considered WR+O star binaries suggest that most 
of them did undergo mass transfer in the past, which stripped the 
hydrogen-rich envelope of the donor stars and the O star companions may 
have rejuvenated due to accretion \citep{braun1995}. The expected masses of the 
WR stars at core helium depletion were calculated using the evolutionary 
tracks of hydrogen deficient, post-Roche Lobe overflow, core helium 
burning star models of \citet{vanbeveren1998a}. For a WR star 
of the nitrogen sequence (i.e. WN star), the WR star was assumed to be at 
the beginning of the helium burning. On the other hand, if a WR star 
was of the carbon sequence (WC star), the calculation was started from the 
point during core helium burning when helium burning products appear at 
the stellar surface due to wind mass loss. This assumption neither affects 
the main results of V20 nor this study (see appendix of V20 for a discussion).
Following this evolution, the expected mass of the WR star at the end of 
core helium burning was calculated. The orbital periods of the WR+O binaries 
at the end of core helium burning of the WR stars were estimated using the 
close binary evolutionary models of \citet{vanbeveren1998b}.

At the end of core helium depletion, we assume that the WR stars will 
directly collapse into BHs of the same mass without any natal kick. 
It means that we do not account for the binary disruption which might 
be induced by high natal kicks. We also neglect the changes in orbital 
separation and eccentricity provoked by natal kicks. We thus expect the 
number of wind-fed BH HMXBs predicted from our analysis to be an upper 
limit on the actual number. Below, we test this assumption  a posteriori 
by comparing our predicted number of wind-fed BH HMXBs with 
observations. We note that a small natal kick may not lead to 
disruption of the binary but introduce an eccentricity in the 
orbit that may result in the production of X-ray at periastron passage. 
In such a case, the X-ray emission is expected to be periodic and 
active only for a small fraction of the orbital period. Therefore, 
we do not expect a small natal kick to significantly alter our results. 

\subsubsection{The BH+O phase}

After the formation of the BH, the orbital evolution is driven by the 
mass loss from the O star companion, which reduces the mass of the O 
star and carries away orbital angular momentum \citep{quast2019,IEM2020}. 
Whether the orbit shrinks or expands depends on the mass ratio and the 
fraction of wind material escaping from the system \citep[see fig. 10 in][]{IEM2020}. 
In our case, the ratio of O star masses to BH masses are below 5, and more than $\sim$99\% 
of wind material escapes from the binary (see Fig.\,\ref{mdot_acc}). 
This implies that we can assume that the orbital parameters remain 
unchanged during the BH+O phase. Considering the fact that most of 
these systems might have undergone a mass transfer episode in the 
past, we also assume that the orbit is circular. 

We follow the subsequent evolution of the O star companions 
in the BH+O binaries by interpolating in the massive single star 
models of \citet{ekstrom2012}. Due to past mass transfer from the 
WR progenitors to the O star companions, the O stars can be found 
to be younger than the age of the binaries, by the process 
of rejuvenation \citep{braun1995}. This is the so-called rejuvenated 
ages of the O stars. The rejuvenated ages of the O stars were obtained 
by V20 from their observed mass, spectral type and luminosity class. 
Here, we estimated the rejuvenated ages of the O stars at the time of 
BH formation by reproducing V20's results with their assumptions. For 
the systems not expected to become detectable BH+O binaries by V20, 
the rejuvenated ages of the O stars at the time of BH formation are 
set to be zero. This does not affect our results as we also find no 
X-ray emission from those systems during the BH+O phase. We assume 
that the BH+O phase lasts until the O stars leave the main sequence 
or fill their Roche lobes, whichever is earlier. On the other hand, 
V20 assumed that the BH+O phase lasts until the O stars fill its Roche 
lobes. 

\subsection{Wind-captured disks during the BH+O phase}
\label{wind-captured_disks}
Due to the gravitational field of the BH, a fraction of the stellar 
wind from the O star can be captured by the BH \citep[][]{Illarionov1975}. 
As a result, a wind-captured disk may form around the BH \citep{shapiro1976,iben1996}. 
Due to turbulent viscosity produced by instabilities such as the 
magneto-rotational instability \citep{Balbus1991}, accreting 
material moves inward in an optically thick and geometrically thin 
accretion disk, in which gravitational energy is efficiently 
converted into thermal energy, producing X-ray emission \citep{Shakura1973}. 

\subsubsection{Wind velocity}
\label{wind_velocity}

The O star wind velocity ($\upsilon_{\rm wind}$) at the location of the BH 
can be approximated as 
\begin{equation}
    \upsilon_{\rm wind} = \upsilon_{\infty}\,\left(1-\frac{R_{\rm O}}{a}\right)^{\beta},
    \label{vwind}
\end{equation}
where $a$ is the orbital separation, $\upsilon_{\infty}$ is the terminal 
velocity of stellar wind and $R_{\rm O}$ is the radius of the O star. For 
O stars (effective temperature higher than 30 kK), the value of $\beta$ 
is 0.8 - 1 \citep{groenewegen1989,puls1996} and the terminal velocity is 
given by \citep{vink2001} 
\begin{equation}
    \upsilon_\infty \simeq 2.6\, \upsilon_{\rm esc},
\end{equation}
where $\upsilon_{\rm esc}$ is the modified escape velocity of O star 
\begin{equation}
    \upsilon_{\rm esc} = \sqrt{\frac{2GM_{\rm O}}{R_{\rm O}}\left(1 - \Gamma\right)}\,,
    \label{vesc}
\end{equation}
$\Gamma$ is the Eddington factor, and $M_{\rm O}$ is the mass of the O star companion.

\subsubsection{Disk formation}
\label{disk_formation_condition_subsection}
A necessary condition for the formation of a wind-captured disk around a BH is 
\begin{equation}
    \frac{R_{\rm disk}}{R_{\rm ISCO}} > 1,
    \label{disk-formation}
\end{equation}
where $R_{\rm ISCO}$ is the radius of the innermost stable orbit and $R_{\rm disk}$ 
is the circularisation radius of a Keplerian accretion disk, defined by
\begin{equation}
    R_{\rm disk} = \frac{j^2}{GM_{\rm BH}},
    \label{rdisk}
\end{equation}
where $j$ is the specific angular momentum of the captured wind material, 
$G$ is the gravitational constant, and $M_{\rm BH}$ is the mass of the BH. 
The radius of the innermost circular orbit around a BH is evaluated by 
\begin{equation}
    R_{\rm ISCO} = \frac{6GM_{\rm BH}}{c^2}\gamma_\pm,
    \label{risco}
\end{equation}
where $c$ is the speed of light, $\gamma_\pm$ represents the modification 
caused by the BH spin with respect to the disk on the location of the 
innermost stable circular orbit. It ranges from 1/6 for a maximally rotating 
BH surrounded by a prograde disk to 3/2 for a maximally rotating BH 
surrounded by a retrograde disk, assuming the disk and BH angular 
momenta are aligned \citep{elmellahthesis}. For a non-rotating BH, 
$\gamma_\pm=1$. \citet{qin2018} found that the spin of the first formed 
BH in a binary is usually very low. 
For a considerable change of the spin, the BH needs to accrete an amount of 
mass of the order of its own mass \citep{wong2012}. Regardless of the birth 
spin of the BH, we assume that the spin of the BH does not change during the 
BH+O phase as only a small fraction of the BH mass is accreted during this phase. 

The specific angular momentum ($j$) accreted by the BH from the O star wind 
can be written as \citep[][eq. 7]{shapiro1976}
\begin{equation}
    j = \frac{1}{2} \eta \Omega_{\rm orb} R_{\rm acc}^2,
    \label{j_SL}
\end{equation}
where $\Omega_{\rm orb}$ is the orbital angular velocity, $\eta$ is a numerical 
factor which quantifies the efficiency of specific angular momentum accretion 
by the BH from the available wind matter and $R_{\rm acc}$ is the accretion 
radius which is the typical distance to the BH at which the wind trajectory 
and/or speed is significantly altered by the gravitational field of the BH. 
It can be written as \citep{davidson1973} 
\begin{equation}
R_{\rm acc} = \frac{2GM_{\rm BH}}{\upsilon_{\rm rel}^{2}},
\label{R_acc}
\end{equation}
where $\upsilon_{\rm rel}=\sqrt{\upsilon_{\rm wind}^2 + \upsilon_{\rm orb}^2}$ 
is the relative velocity of the stellar wind with respect to the BH for a 
circular orbit, $\upsilon_{\rm wind}$ is the wind velocity of the O star companion and 
$\upsilon_{\rm orb}$ is the relative velocity of the BH with respect to the O star, that is, 
$\upsilon_{\rm orb}$ = $\Omega_{\rm orb}a$. 

Eq.\,\eqref{j_SL} was obtained under the assumption that the wind velocity is 
considerably larger than the orbital velocity, which is consistent with our 
further analysis (see Fig.\,\ref{vw-vorb}). If all wind material entering the 
accretion radius can be accreted by the BH, $\eta = 1$ \citep{shapiro1976}. 
Detailed hydrodynamical simulations suggest that this efficiency factor can 
be lower, $\sim$1/3 \citep{livio1986,ruffert1999}. In what follows, we consider 
these two values. 

Defining the mass ratio $q = M_{\rm O}/M_{\rm BH}$ and combining Eqs.\,\eqref{rdisk} - \eqref{R_acc}, 
the disk formation criterion can be converted into the dimensionless form 
\begin{equation}
  \frac{2}{3}\frac{\eta^2}{(1+q)^2} > \left(\frac{\upsilon_{\rm orb}}{c}\right)^2\left(1+\frac{\upsilon_{\rm wind}^2}{\upsilon_{\rm orb}^2}\right)^4\gamma_\pm,
  \label{disk-formation-2}
\end{equation}
or equivalently
\begin{equation}
    \frac{\rd}{\risco} =  \frac{2}{3}\frac{\eta^2}{(1+q)^2} \left(\frac{\upsilon_{\rm orb}}{c}\right)^{-2}\left(1+\frac{\upsilon_{\rm wind}^2}{\upsilon_{\rm orb}^2}\right)^{-4}\gamma_\pm^{-1} > 1.
    \label{disk-formation-3}
\end{equation}
Eqs.\,\eqref{disk-formation-2} and \eqref{disk-formation-3} suggest 
that a wind-captured disk can form around a BH if the captured 
material carries enough angular momentum, the wind speed is low 
compared to the orbital speed, and the orbital speed is high. 

\subsection{X-ray luminosity}

\label{xray_lum_subsection}

We can distinguish three cases for the morphology of the accretion flow: sub-Eddington 
accretion via a disk, super-Eddington accretion via a disk and spherical 
accretion. The first two happen only if enough angular momentum is carried 
by the accretion flow (see Sect.\,\ref{wind-captured_disks}). Super-Eddington 
accretion occurs when the mass accretion rate is so high that the X-ray 
luminosity it produces exceeds the Eddington luminosity of the BH. 
Although super-Eddington accretion onto neutron stars has been observed in 
ultra-luminous X-ray sources \citep{Bachetti2014,Fuerst2016,Israel2017}, the 
typical mass accretion rate calculated in our study is much 
smaller than the Eddington accretion rate for the individual systems 
(Fig.\,\ref{mdot_acc}). Accretion disks with a sub-Eddington mass accretion 
rate are thought to be geometrically thin and optically thick 
centrifugally-maintained structures \citep{Shakura1973,Novikov1973}. 
Notwithstanding minor relativistic corrections, such a disk around a BH 
radiates mostly in X-rays, and the maximum associated luminosity 
is \citep{FKR2002,elmellahthesis}:
\begin{equation}
    L_{\rm X}=\frac{1}{2}\frac{GM_{\rm BH}\dot{M}_{\rm acc}}{R_{\rm ISCO}},
    \label{Lx_eqn}
\end{equation}
where $\dot{M}_{\rm acc}$ is the mass accretion rate. 

In order to evaluate the mass accretion rate, we rely on the wind accretion 
formula introduced by \citet{davidson1973} \citep[see also the review by][]{Edgar2004}. 
It is valid in binary systems provided the wind speed at the binary orbital 
separation is larger than the orbital speed \citep{elmellah2017}. In this 
case, the fraction of the accreted wind can be approximated by 
\begin{equation}
    \frac{\dot{M}_{\rm acc}}{\dot{M}_{\rm wind}} = \frac{1}{4} \left(\frac{R_{\rm acc}}{a}\right)^2 \frac{\upsilon_{\rm rel}}{\upsilon_{\rm wind}},
\end{equation}
where $\dot{M}_{\rm wind}$ is the O star wind mass loss rate. 

Finally, in the case of spherical accretion, the mass accretion rate 
is not an independent variable. Instead, it is set by the location of 
the sonic point as described in the 1D spherical Bondi model \citep{Bondi1952}. 
Without an accretion disk, thermal bremsstrahlung dominates the radiation 
from the optically thin wind material, which makes spherical accretion 
radiatively inefficient \citep{shapiro1983}. We do not expect this regime to 
produce any X-ray emission above detectable levels.

\subsection{detectability of a BH+O system}

The X-ray active lifetime ($\tau_{\rm Lx}$) of each BH+O binary model 
considered is defined as the amount of time during the BH+O phase when 
the system is detectable as a wind-fed BH HMXB. We assume that this 
will be the case only when an accretion disk forms, that is, Eq.\,\eqref{disk-formation-3} 
is satisfied, and when the calculated X-ray luminosity (Eq.\,\ref{Lx_eqn}) 
and the distance to the source yield a flux above a detection threshold 
that we set to $\sim$10$^{-11}$ erg s$^{-1}$ cm$^{-2}$. Our adopted 
threshold is similar to the flux detection limit of non-focussing X-ray 
telescopes with typical integration times \citep{wood1984,bradt1991,zand1994}. 
We discuss the relevance of the X-ray flux threshold in the 
light of the sensitivity of current all-sky monitoring X-ray instruments 
in Sect.\,\ref{dnc}. 

\section{Results}
\label{results}

\begin{table*}[t]
\centering
\caption{Predicted X-ray active lifetime ($\tau_{\rm Lx}$, in Myrs) of each of the 17 BH+O binary models and expected number of wind-fed black hole high mass X-ray binaries ($N_{\rm XRBs}$, last line), for various combinations of $\beta$, $\eta$ and $\gamma_{\pm}$. The bold highlighted column represents our fiducial case.}
\begin{tabular}{c|ccc|ccc|ccc|ccc}
\hline\hline
&  \multicolumn{3}{|c|}{$(\beta,\eta)$\tablefootmark{a}$ = (1, 1)$} & \multicolumn{3}{c|}{$(\beta,\eta) = (0.8, 1)$} & \multicolumn{3}{c|}{$(\beta,\eta) = (1, 1/3)$} & \multicolumn{3}{c}{$(\beta,\eta) = (0.8, 1/3)$} \\
$\gamma_\pm$ & 1/6 & 1 & 3/2 & 1/6 & 1 & 3/2 & 1/6 & 1 & 3/2 & 1/6 & 1 & 3/2 \\
\hline
WR155	&	1.5	&	1.5	&	1.4	&	1.5	&	1.2	&	0.9	&1.4	&	\textbf{0.2}	&	0.0	&	0.9	&	0.0	&	0.0\\
WR151	&	3.6	&	1.6	&	1.3	&	2.9	&	1.1	&	0.8	&1.3	&	\textbf{0.3}	&	0.1	&	0.8	&	0.0	&	0.0\\
WR139	&	2.0	&	1.0	&	0.8	&	1.6	&	0.6	&	0.5	&0.8	&	\textbf{0.2}	&	0.1	&	0.5	&	0.0	&	0.0\\
WR31	&	1.9	&	1.0	&	0.8	&	1.6	&	0.7	&	0.6	&0.8	&	\textbf{0.2}	&	0.1	&	0.6	&	0.0	&	0.0\\
WR42	&	1.8	&	0.8	&	0.7	&	1.7	&	0.6	&	0.5	&0.7	&	\textbf{0.0}	&	0.0	&	0.5	&	0.0	&	0.0\\
WR47	&	0.6	&	0.4	&	0.4	&	0.6	&	0.4	&	0.3	&0.4	&	\textbf{0.1}	&	0.0	&	0.3	&	0.0	&	0.0\\
WR79	&	0.8	&	0.1	&	0.0	&	0.6	&	0.0	&	0.0	&0.0	&	\textbf{0.0}	&	0.0	&	0.0	&	0.0	&	0.0\\
WR127	&	0.6	&	0.0	&	0.0	&	0.4	&	0.0	&	0.0	&0.0	&	\textbf{0.0}	&	0.0	&	0.0	&	0.0	&	0.0\\
WR21	&	0.3	&	0.2	&	0.1	&	0.3	&	0.1	&	0.1	&0.1	&	\textbf{0.0}	&	0.0	&	0.1	&	0.0	&	0.0\\
WR9	&	0.1	&	0.0	&	0.0	&	0.1	&	0.0	&	0.0	&0.0	&	\textbf{0.0}	&	0.0	&	0.0	&	0.0	&	0.0\\
WR97	&	0.0	&	0.0	&	0.0	&	0.0	&	0.0	&	0.0	&0.0	&	\textbf{0.0}	&	0.0	&	0.0	&	0.0	&	0.0\\
WR30	&	0.0	&	0.0	&	0.0	&	0.0	&	0.0	&	0.0	&0.0	&	\textbf{0.0}	&	0.0	&	0.0	&	0.0	&	0.0\\
WR113	&	0.0	&	0.0	&	0.0	&	0.0	&	0.0	&	0.0	&0.0	&	\textbf{0.0}	&	0.0	&	0.0	&	0.0	&	0.0\\
WR141	&	0.0	&	0.0	&	0.0	&	0.0	&	0.0	&	0.0	&0.0	&	\textbf{0.0}	&	0.0	&	0.0	&	0.0	&	0.0\\
WR35a	&	0.0	&	0.0	&	0.0	&	0.0	&	0.0	&	0.0	&0.0	&	\textbf{0.0}	&	0.0	&	0.0	&	0.0	&	0.0\\
WR11	&	0.0	&	0.0	&	0.0	&	0.0	&	0.0	&	0.0	&0.0	&	\textbf{0.0}	&	0.0	&	0.0	&	0.0	&	0.0\\
WR133	&	0.0	&	0.0	&	0.0	&	0.0	&	0.0	&	0.0	&0.0	&	\textbf{0.0}	&	0.0	&	0.0	&	0.0	&	0.0\\
\hline
$N_{\rm XRBs}$\tablefootmark{b}&	33.0	&	16.6	&	13.4	&	28.3	&	11.8	&	9.0	&13.4	&	2.5	&	0.7	&	9.0	&	0.0	&	0.0\\
\hline
\end{tabular}

\tablefoot{
\tablefoottext{a}{see Eqs.\,\eqref{vwind} and \eqref{j_SL} for the definition of $\beta$ and $\eta$ respectively.}
\tablefoottext{b}{$N_{\rm XRBs}$ is the predicted number of wind-fed BH high mass X-ray binaries by taking 0.4 Myr as the typical lifetime of WR stars. See Sect.\,\ref{results} for more details.}
}
\label{main_table}
\end{table*}

\subsection{Fiducial parameter set}

\begin{figure}
    \centering
    \includegraphics[width=\linewidth]{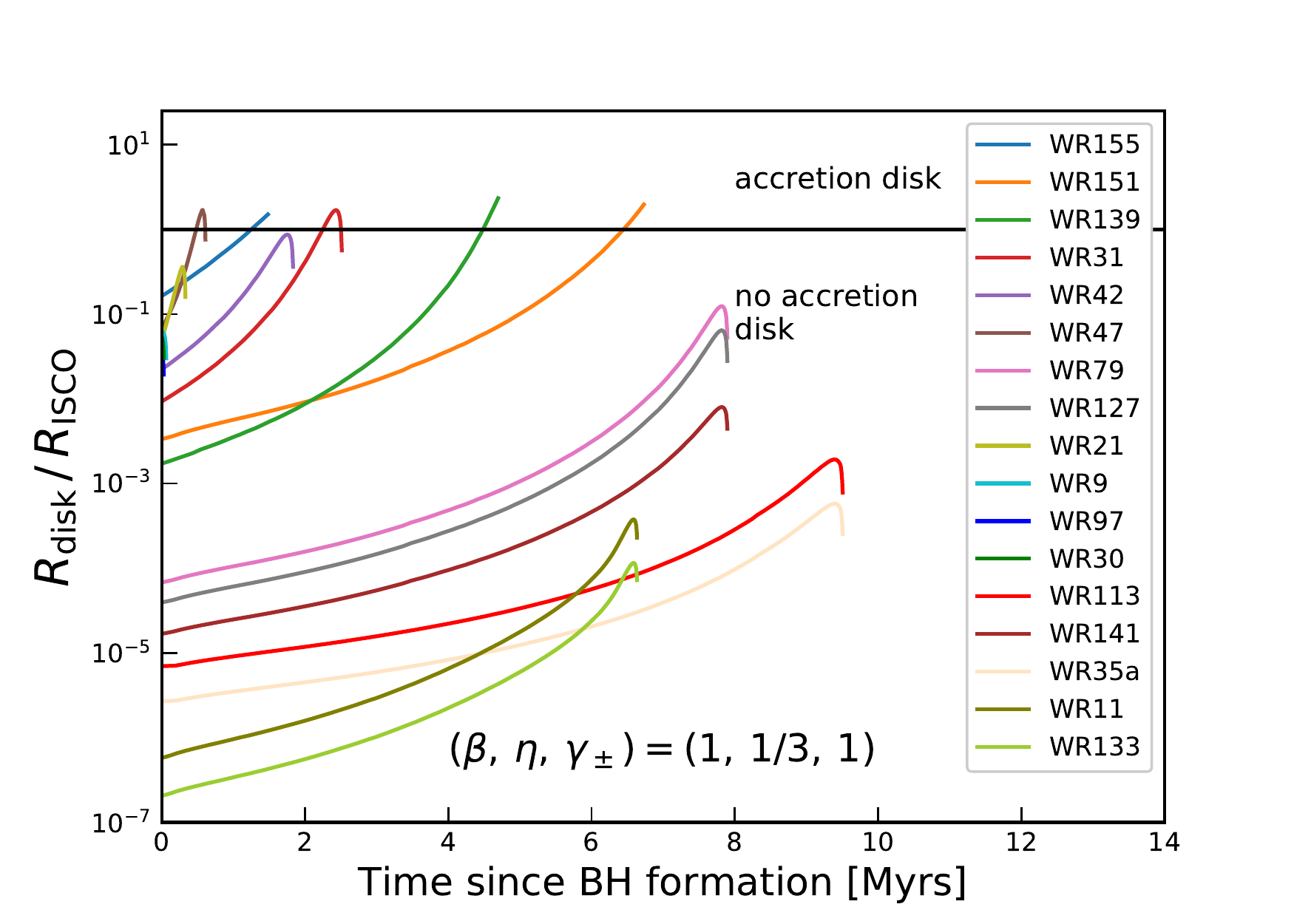}
    \caption{Evolution of the ratio of circularisation radius $R_{\rm disk}$ to the radius of innermost stable circular orbit $R_{\rm ISCO}$ during the BH+O phase as a function of the time since the formation of the BH, for $(\beta,\,\eta,\,\gamma_\pm)=(1,\,1/3,\,1)$. The black horizontal line shows the dividing line above which an accretion disk is expected. The colour coding in the legend identifies the 17 progenitor WR+O star systems that are expected to give rise to the BH+O binaries.}
    \label{plot_result}
\end{figure}

\begin{figure}
    \centering
    \includegraphics[width=\linewidth]{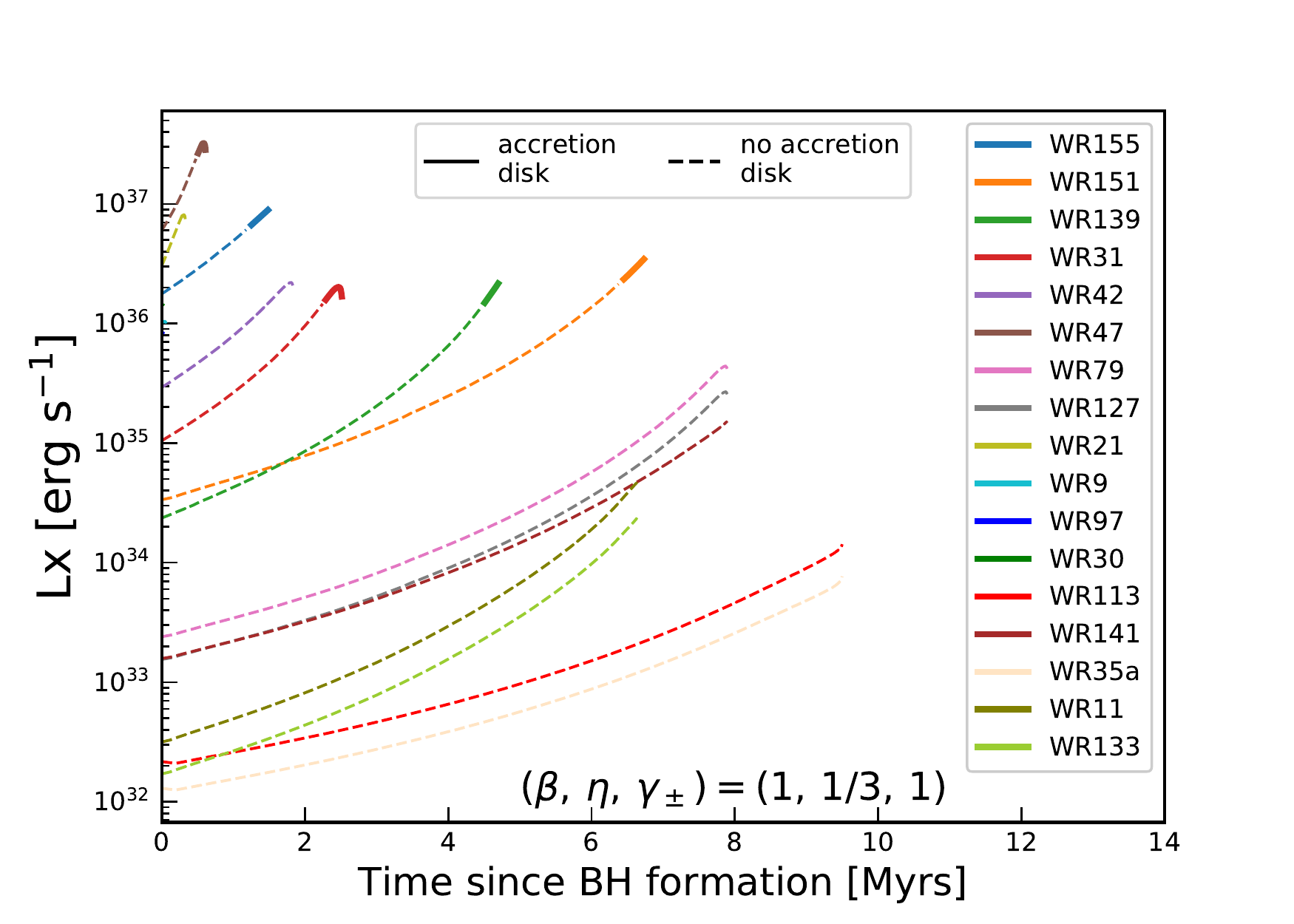}
    \includegraphics[width=\linewidth]{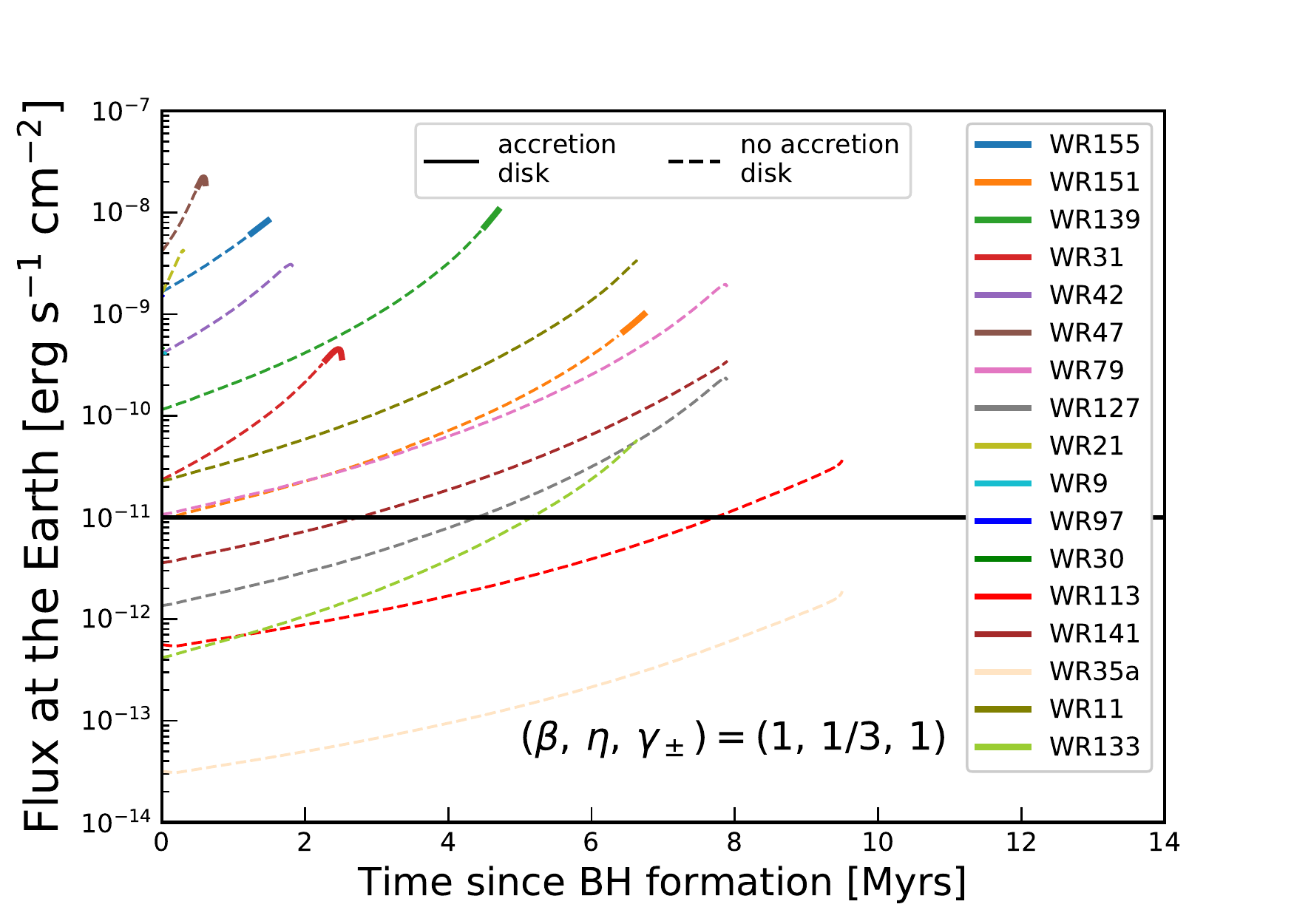}
    \caption{Evolution of: X-ray luminosity (\textit{upper panel}) calculated using Eq.\,\eqref{Lx_eqn}, and the corresponding X-ray flux at Earth (\textit{lower panel}) for our BH+O models when an accretion disk can form according to our criterion (solid line). The dashed lines indicate the X-ray luminosity and flux evolution if an accretion disk could form for the entire BH+O phase. The black horizontal line shows our adopted flux detection limit.}
    \label{fiducial_lum_flux}
\end{figure}

Figure \ref{plot_result} shows the evolution of the ratio of 
circularisation radius ($R_{\rm disk}$) to the radius of innermost 
stable circular orbit ($R_{\rm ISCO}$) during the BH+O phase, for 
the 17 progenitor WR+O star binaries. For our fiducial case, we 
adopt the value of $\beta$=1 \citep{vink2001}, $\eta$=1/3 \citep{elmellahthesis} 
and $\gamma_\pm$=1 \citep{qin2018}. The O star expands during 
core hydrogen burning, leading to a decrease in its wind velocity, 
which makes the formation of a wind-captured disk easier during 
the late stages of its main sequence evolution. In most systems, 
there is a small decrease in the ratio of the circularisation 
radius to the innermost stable circular orbit towards the end 
of the BH+O star phase, which is related to the shrinkage of 
massive stars when they approach their core hydrogen depletion. 
WR139, WR151, and WR155 do not present this feature since their 
BH+O phases are terminated due to the Roche Lobe filling condition 
before their O stars complete core hydrogen burning. 
Whereas the mass ratio of WR\,139 suggests this system will 
merge at this time, the other two could undergo an SS433-like 
evolution leading to short-period WR+BH binaries \citep{Heuvel2017}, 
which lies outside our present scope.

We find that no accretion disk forms in 12 of our BH+O models. Among 
them, three systems are not visible in this plot since the estimated rejuvenated 
age of their O stars are very close to the O stars' main sequence lifetime, 
such that the duration of their BH+O phase are very small. Importantly, 
we find that only in 5 BH+O models, all with orbital periods $\leq$10 days, an 
accretion disk can form for a small fraction of the total the BH+O phase. For 
systems with higher orbital periods, an accretion disk does not form at all for 
the entire BH+O phase. Noting that the orbital period of WR 22 is $\sim$80 days, 
we do not expect that the BH+O binary anticipated to form from WR 22 will be 
X-ray bright at any time.

Figure \ref{fiducial_lum_flux} shows the X-ray luminosity (top panel) and its corresponding 
flux at Earth (bottom panel), calculated using Eq.\,\eqref{Lx_eqn} for our BH+O models. We find 
that when an accretion disk can form, the predicted X-ray flux at Earth is well 
above the flux detection limit we have assumed. In other words, the X-ray luminosity 
from the accretion disk is not a bottleneck for our BH+O models to be detectable in 
X-rays. We note that Eq.\,\eqref{Lx_eqn} only holds when an accretion disk is present 
such that the dashed lines are only indicative of the X-ray luminosity and flux if 
an accretion disk could form for the entire BH+O phase. The X-ray luminosity from a BH+O system 
without an accretion disk is expected to be orders of magnitude lower than what is 
predicted by Eq.\,\eqref{Lx_eqn} (see discussion in Sect.\,\ref{xray_lum_subsection}). 

For each system where an accretion disk can form, we calculate the duration 
for which it will detectable as an wind-fed BH HMXB (i.e. 
the X-ray active lifetime). To predict the number 
of wind-fed BH HMXB systems we expect based on the 17 progenitor WR+O star 
systems, we assume (as in V20) that the observed numbers of WR+O binaries 
and wind-fed BH HMXBs are proportional to the lifetime in the respective 
phases. One WR+O binary is thus representative of $\tau_{\rm Lx}/\tau_{\rm WR}$ 
wind-fed BH HMXBs, where $\tau_{\rm WR}$ is the duration of the WR+O binary phase. 
Taking $\tau_{\rm WR} =$ 0.4 Myrs as the typical lifetime of a WR star (V20), 
we expect $\sim$2.5 wind-fed BH HMXBs from the 17 WR+O binaries. Accounting for 
the observational and theoretical bias in the population of WR+O binaries (see 
discussion in Sect.\,\ref{sample}), it is likely that the number of wind-fed BH 
HMXBs in the entire Milky Way would be $\sim$2-3.

\subsection{Effects of parameter variations}

The predicted number of wind-fed BH HMXBs is sensitive to the uncertainties in 
the parameters we have assumed. We explore the results computed using reasonable 
variations to our fiducial parameter set in Table\,\ref{main_table}. For a non-
rotating BH ($\gamma_{\pm}$=1), by varying ($\beta$,\,$\eta$) from (0.8, 1/6) to 
(1, 1/2), the predicted number of wind-fed BH HMXBs out of 17 WR+O binaries varies 
from 0 to 16.6. Considering a maximally spinning BH with a prograde accretion 
disk, the predicted number can be boosted up to 33, suggesting an 
observational bias in favour of wind-fed BH HMXBs containing maximally rotating BHs 
surrounded by a prograde disk. In the following subsections, we discuss the 
effects of these parameters individually.

\begin{figure}
    \centering
    \includegraphics[width=\linewidth]{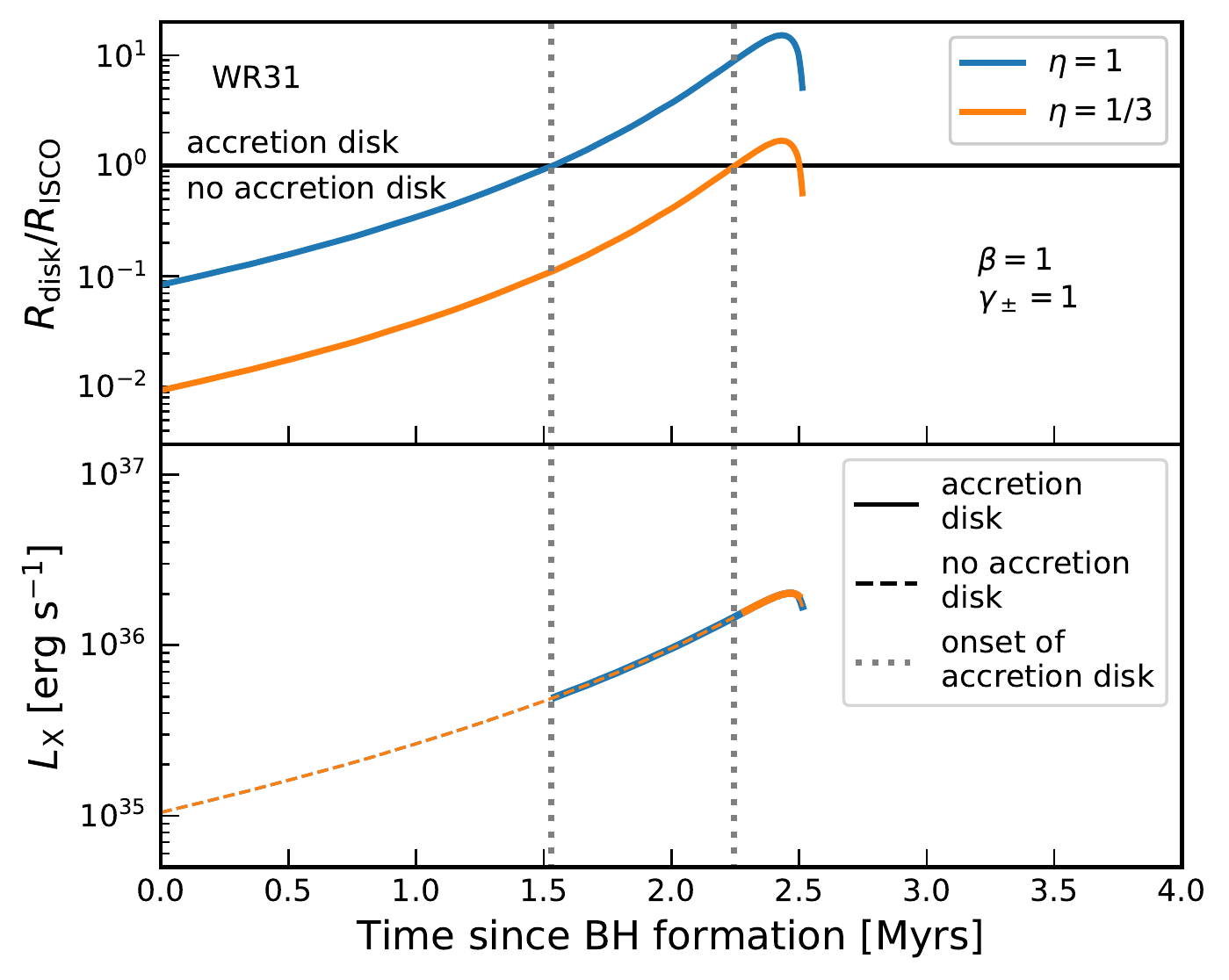}
    \caption{Effects of $\eta$ parameter on the ratio of the circularisation radius to the radius of the innermost stable circular orbit (upper panel) and the X-ray luminosity (lower panel). We take the BH+O binary derived from WR31 as an example. The $\eta$ parameter is taken to be 1 and 1/3 (colour coded), and $(\beta,\,\gamma_\pm) = (1,\,1)$. The line styles in the lower panel have the same meaning as in Fig.\,\ref{fiducial_lum_flux}.}
    \label{eta}
\end{figure}

\subsubsection{Efficiency of specific angular momentum accretion}

From Eq.\,\eqref{disk-formation-3}, the ratio of circularisation 
radius to the radius of the innermost stable orbit varies with 
the square of the efficiency of specific angular momentum 
accretion
\begin{equation}
    \frac{\rd}{\risco} \propto \eta^2.
\end{equation}
The predicted X-ray luminosity when an accretion disk can form does 
not depend on the efficiency parameter. So, the likelihood of the formation 
of an accretion disk in our BH+O models increases with the increase in 
the efficiency of angular momentum accretion by the BH. In Fig.\,\ref{eta}, 
we show the variation of the above two quantities with the efficiency of 
specific angular momentum accretion, for the BH+O model corresponding to 
WR 31. We find that the amount of time an accretion disk can form during 
the BH+O phase is significantly longer when the efficiency of angular 
momentum accretion increases by a factor of 3. On the other hand, the X-ray 
luminosity predicted from Eq.\,\eqref{Lx_eqn} is unaffected. From Table 
\ref{main_table}, we find that the number of predicted wind-fed BH 
HMXBs increases by 6.5 times when the $\eta$ increases from 1/3 to 
1, and the other two parameters are at their fiducial value. 

\subsubsection{Black hole spin}

\begin{figure}
    \centering
    \includegraphics[width=\linewidth]{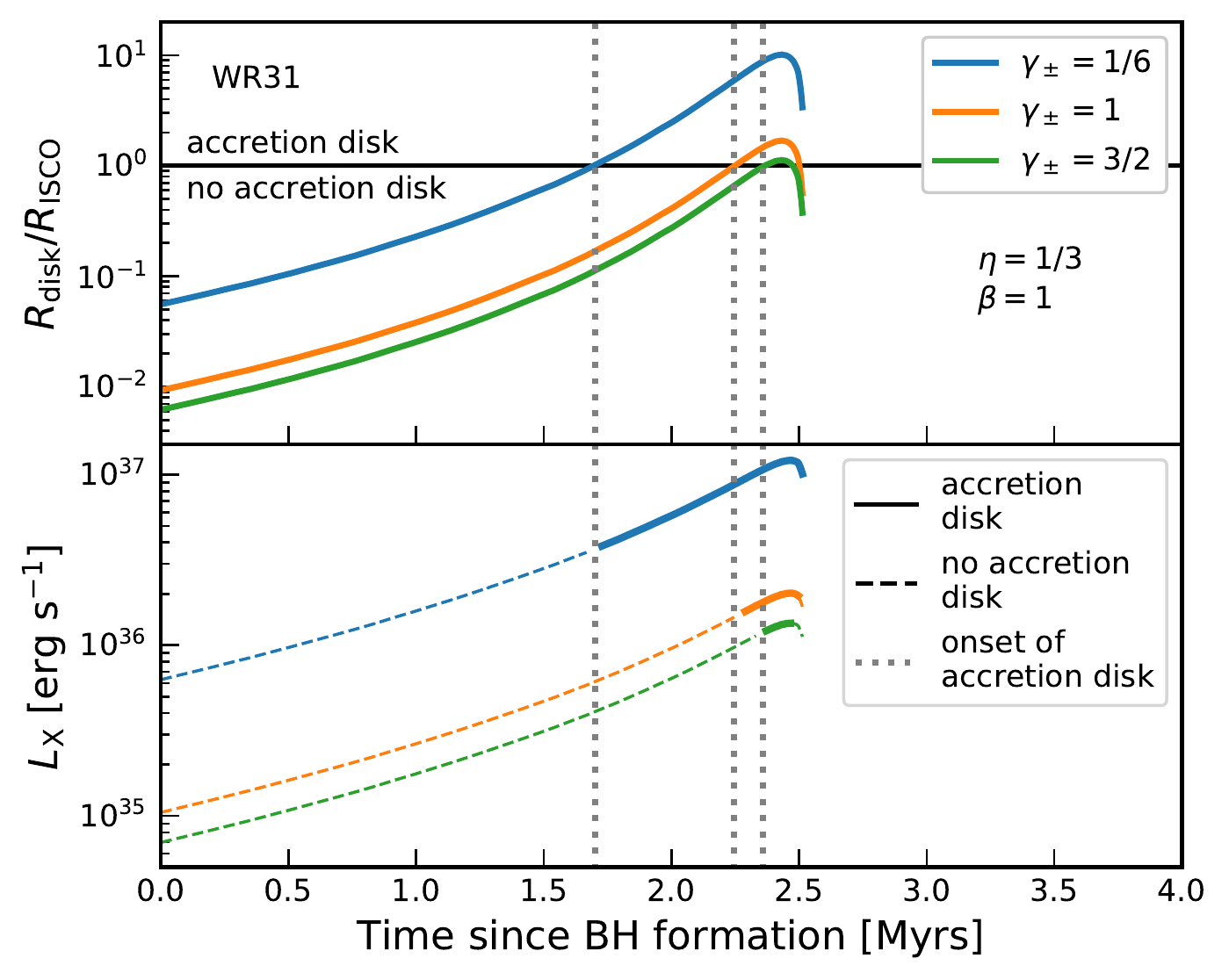}
    \caption{Effects of the $\gamma_\pm$ parameter. The same as Fig.\,\ref{eta} but $\gamma_\pm$ is taken to be 1/6, 1, and 3/2 (colour coded), and $(\beta,\,\eta) = (1,\,1/3)$. The line styles in the lower panel have the same meaning as in Fig.\,\ref{fiducial_lum_flux}.
    }
    \label{gamma}
\end{figure}

Our definition of the radius of innermost stable circular orbit around 
a BH (Eq.\,\ref{risco}) accounts for the effect of the spin of the BH 
on the formation of an accretion disk (via $\gamma_{\pm}$). The spin parameter of 
BHs in observed wind-fed HMXBs can be quite high, as in Cyg X-1 
\citep{gou2011,Zhao2021,miller-jones2021}. To account for the spin 
of the BH, we calculate the predicted number of wind-fed BH HMXB 
derived from the 17 progenitor WR+O binaries for three cases 
(see Table \ref{main_table}): i) When the BHs are maximally rotating 
with a prograde accretion disk, ii) when the BHs are maximally rotating 
with a retrograde accretion disk, and iii) for a non-spinning BH. 

Both the ratio of circularisation radius to the radius of the 
innermost stable orbit, and the X-ray luminosity from an accretion 
disk varies inversely with our BH spin parameter 
\begin{equation}
    \frac{\rd}{\risco} \propto \gamma_\pm^{-1}
\end{equation}
and 
\begin{equation}
    L_{\rm X} \propto \gamma_\pm^{-1}.
\end{equation}
Figure \ref{gamma} shows the effect of the BH spin on the 
formation of an accretion disk and the emitted X-ray luminosity 
during the BH+O phase of WR 31. For a BH maximally rotating with 
a prograde accretion disk, both the amount of time for which an 
accretion disk can form, and the X-ray luminosity predicted from 
the accretion disk increases significantly. In Table \ref{main_table}, 
we find that the predicted number of wind-fed HMXBs increases by 
a factor of $\sim$5 for the case of a maximally rotating BH with 
a prograde disk, and decreases by a factor of $\sim$3.5 for a maximally 
rotating BH with a retrograde disk, compared to a non-rotating BH, with 
the other two parameters are at their fiducial values. 

The fact that we predict a short X-ray active lifetime for 
non-spinning BHs in BH+O systems, while the only observed wind-fed 
BH HMXB in the Milky Way is known to have high spin parameter, hints 
to the possibility that only BHs that were born with a very high spin 
are likely to be detectable as an X-ray source if they are associated 
with an O star in a close binary configuration. 
\citet{qin2019} showed that high spin BHs can be produced only if the 
efficiency of angular momentum transport in stellar models is reduced. 
As such, BHs with high birth spins might be rare, as are observed 
wind-fed BH HMXBs. 

\subsubsection{O star wind velocity law}

\begin{figure}
    \centering
    \includegraphics[width=\linewidth]{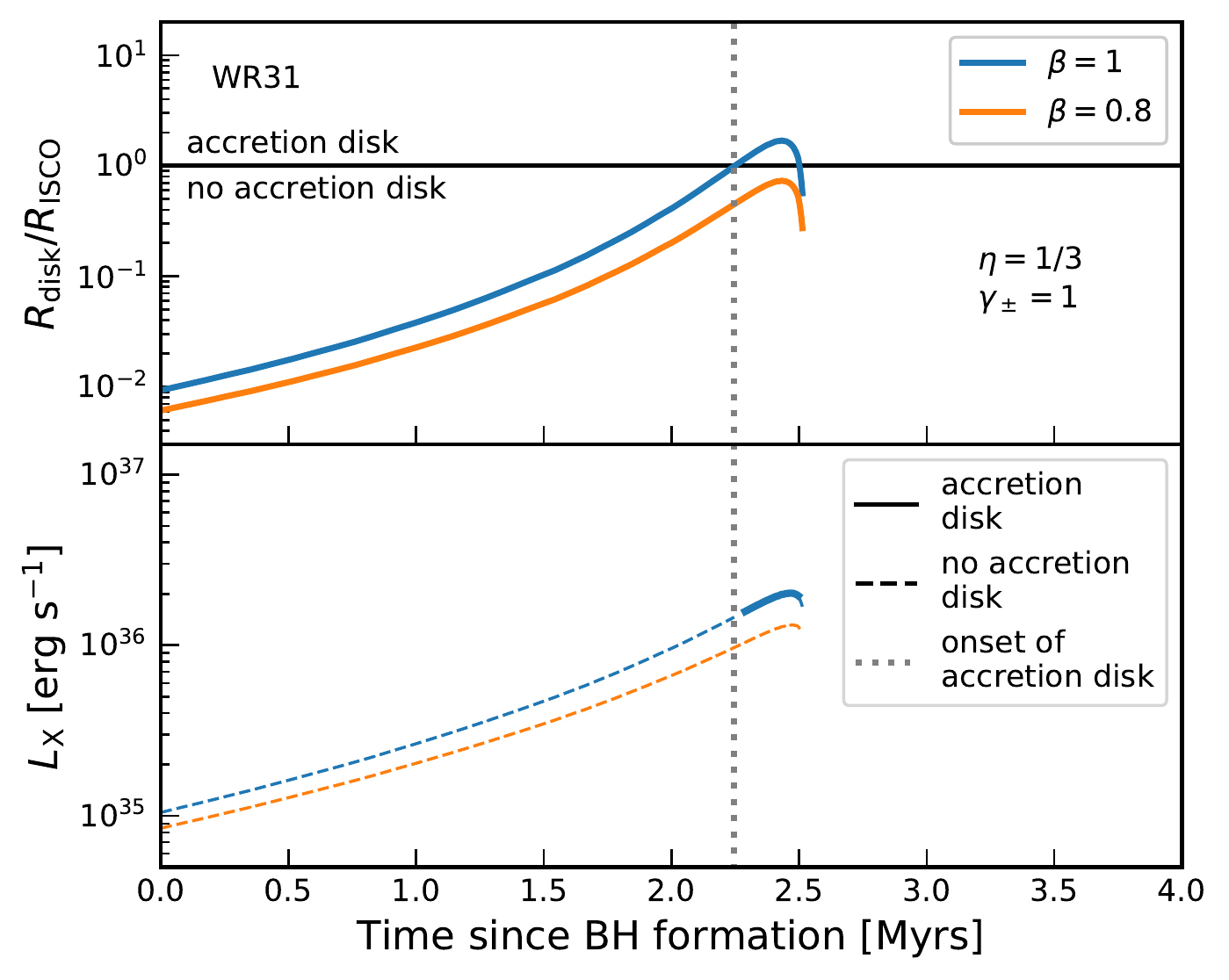}
    \caption{Effects of the $\beta$ parameter. The same as Fig.\,\ref{eta} but $\beta$ parameter is taken to be 0.8 and 1 (colour coded), and $(\eta,\,\gamma_\pm) = (1/3,\,1)$. The line styles in the lower panel have the same meaning as in Fig.\,\ref{fiducial_lum_flux}.
    }
    \label{beta}
\end{figure}

\begin{figure*}
\begin{minipage}{0.5\linewidth}
\includegraphics[width=\linewidth]{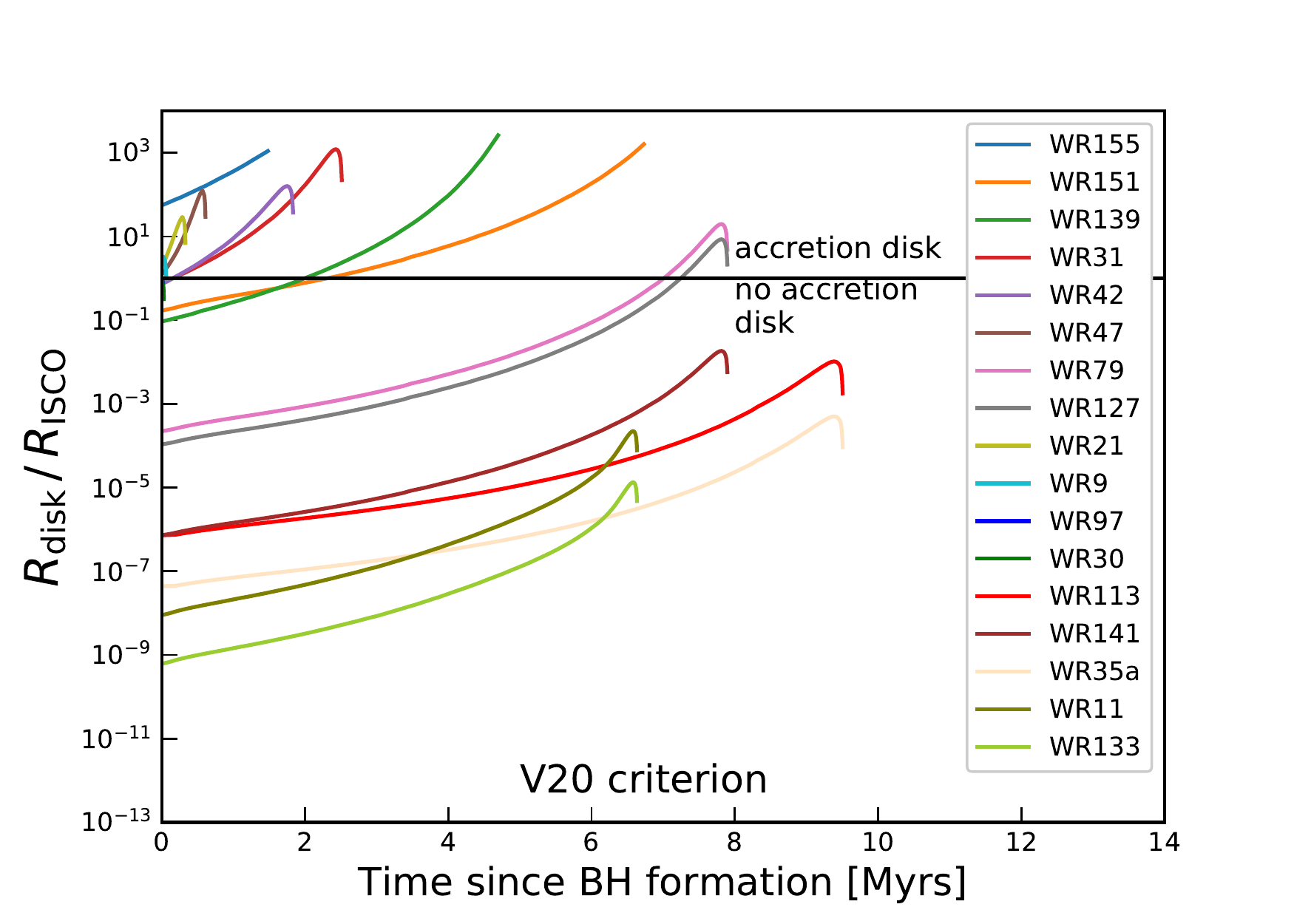}    
\end{minipage}
\begin{minipage}{0.5\linewidth}
\includegraphics[width=\linewidth]{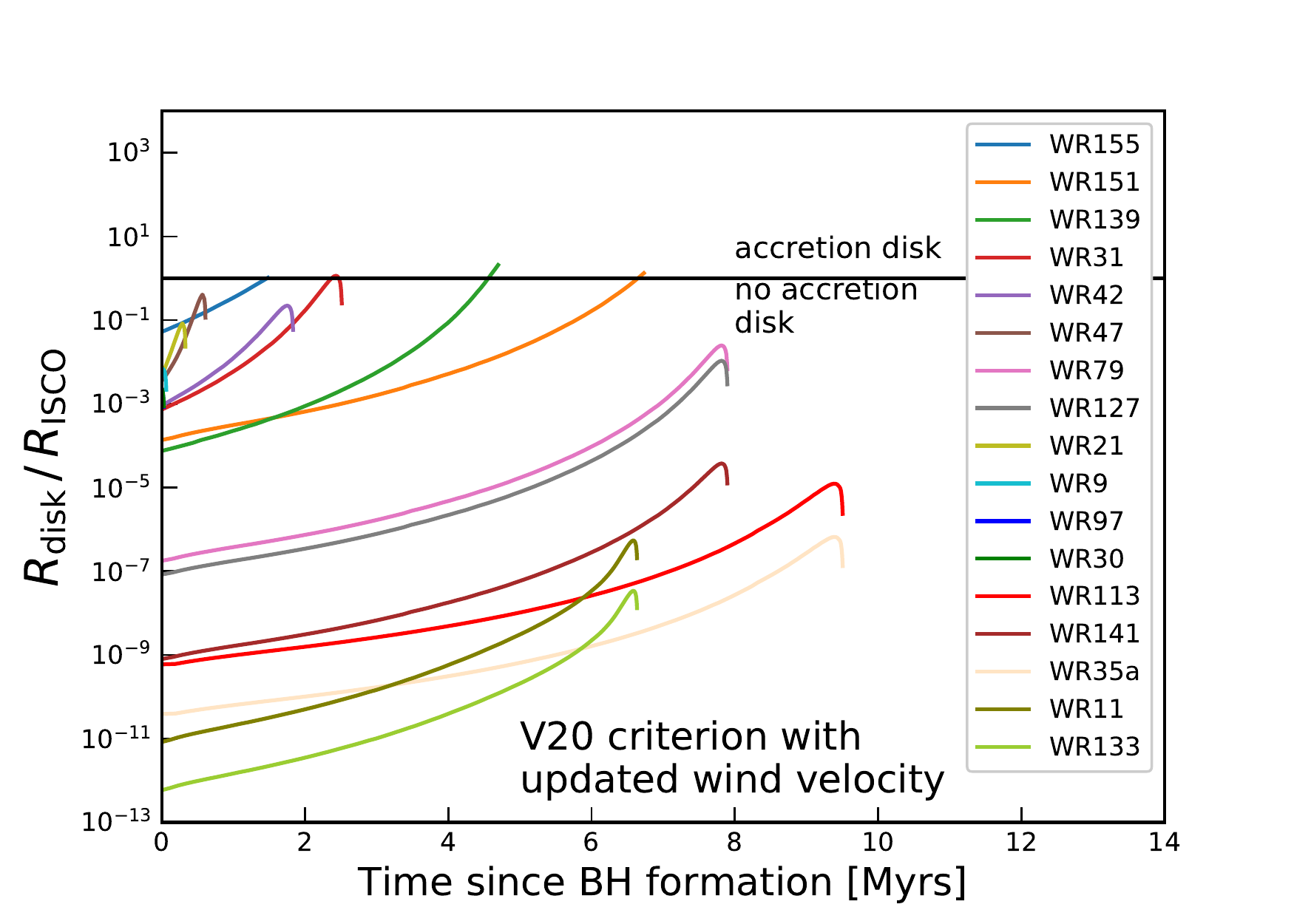}    
\end{minipage}
\caption{Evolution of the ratio of circularisation radius of the accreted wind matter from the O star to the radius of innermost stable circular orbit of the BH for the 17 BH+O star binary models as a function of the time since the formation of the BH. The left panel shows the evolution as calculated by V20, that does not account for the orbital velocity of the companion and the typical wind velocity of O stars. The right panel shows the same evolution when we use the typical O star wind velocity (Eq.\,\ref{vwind}), the mass of the BH, and the orbital velocity of the O star companion. The black horizontal line in both the panels show the dividing line above which an accretion disk can form. The colour coding in the legend identifies the 17 progenitor WR+O star systems that are expected to give rise to the BH+O binaries.
}
\label{fig_compare}
\end{figure*}

The exponent $\beta$ in the wind velocity law for O stars is constrained 
from observations to be 0.8-1 \citep{groenewegen1989,lamers1995,puls1996}. 
Since wind velocity is always larger than orbital velocity in our work, 
Eq.\,\eqref{disk-formation-3} and Eq.\,\eqref{Lx_eqn} suggest the dependencies
\begin{equation}
    \frac{\rd}{\risco} \propto \vw^{-8} \propto \left(1 - \frac{\ro}{a}\right)^{-8\beta},
\end{equation}
and
\begin{equation}
    L_{\rm X}\propto \upsilon_{\rm rel}^{-3} \propto \left(1 - \frac{\ro}{a}\right)^{-3\beta}.
\end{equation}
In our analysis, the ratio between the O star radius 
\citep[obtained from the stellar tracks of ][]{ekstrom2012} and orbital 
separation is generally below 0.4. 
Therefore, changing $\beta$ from 1 to 0.8 can maximally reduce\rdrisco\,by a 
factor of $\sim$2, making the formation of accretion disk more difficult. 
Likewise, the predicted X-ray luminosity from Eq.\,\eqref{Lx_eqn} is also decreased. 

We present the effects of the $\beta$ parameter on the formation of an 
accretion disk and the X-ray luminosity emitted from the disk in Fig.\,\ref{beta}, 
where we take the BH+O model derived from WR 31 as an example. We find 
that even the change in the assumed $\beta$ value from 1 to 0.8 makes the 
BH+O model of WR 31 to become X-ray inactive due to the inability to form 
an accretion disk. The predicted X-ray luminosity and thereby the X-ray 
flux from an accretion disk, if it were to form, is also decreased, but 
not as significantly as to fall beyond our flux detection limit.

From Table \ref{main_table}, we see that when we change the value of the 
$\beta$ from 1 to 0.8 (other parameters remaining at their fiducial 
values), our BH+O binary models do not have any X-ray bright phase. The 
predicted number of wind-fed BH HMXBs decreases from $\sim$2.5 to zero. 
This shows that the X-Ray active lifetime of the BH+O binaries analysed 
in our work is very sensitive to the assumed wind velocity of the O star 
companion.

\section{Comparison with earlier work}
\label{earlier_work}

Starting from the same 17 WR+O binaries, V20 performed a similar analysis 
and predicted to find over 200 wind-fed BH HMXBs in the Milky Way. 
Here, we compare the analysis of V20 with our work and discuss the factors 
that lead to the difference in the predicted numbers.

V20 adopted the accretion disk formation criterion derived by \citet{iben1996}. 
\citet{iben1996} primarily modelled accretion onto degenerate white dwarfs 
from red giant donors and their central idea remained the same 
in the sense that they assumed that an accretion disk forms when the specific 
angular momentum of the accreted matter exceeds that of the innermost 
stable circular orbit radius of the BH. \citet{iben1996} 
assumed that the specific angular momentum ($j$) accreted by the degenerate 
dwarf from the stellar wind of the giant companion is given by 
\begin{equation}
j \sim \Omega_{\rm g} R_{\rm g}^{2} \left( \frac{R_{\rm acc}}{a} \right)^{2},
\label{specific_j_van}
\end{equation}
where $\Omega_{\rm g}$ and $R_{\rm g}$ are the angular velocity and 
radius of the giant star respectively. 
\citet{iben1996} defined the accretion radius $R_{\rm acc}$ as 
\begin{equation}
R_{\rm acc} = \frac{2GM_{\rm dd}}{\upsilon_{\rm wind}^{2}},
\label{R_BH_van}
\end{equation}
where $M_{\rm dd}$ is the mass of the degenerate dwarf. 
They further assumed that the companion star is tidally 
locked and the binary mass is dominated by the giant star mass ($M_{\rm g}$). 
Comparing our work and V20, we note the difference of a factor 
$\sim$R$_{\rm g}^{2}$/a$^2$ in the definition of specific angular 
momentum accretion, and the omission of the relative velocity of the BH 
with respect to the main sequence companion in the definition of the accretion radius. 
For the wind velocity, \citet{iben1996} assume that the wind velocity 
from the giant companions is given by
\begin{equation}
    \upsilon'_{\rm wind} = \upsilon'_{\rm esc}\,\left(1-\frac{R_{\rm g}}{a}\right),
    \label{vw_van}
\end{equation}
where $\upsilon'_{\rm esc}=\sqrt{2GM_{\rm g}/R_{\rm g}}$ is the escape 
velocity from the companion star. Effectively, they assumed that the 
terminal wind velocity is equal to the escape velocity from the surface 
of the star, and $\beta$=1. They also did not take into account the Eddington factor.

Observational studies of the terminal wind velocities of O stars show 
that their terminal velocities are larger than their escape velocities, 
such that the appropriate expression for the wind velocity from O stars 
is given by Eq.\,\eqref{vwind} \citep{vink2001}.
However, V20 did not account for the typical wind velocity of the O stars 
when they adopted the disk formation criterion derived by \citet{iben1996} 
for their BH+O systems, that is, the terminal wind velocities used by V20 in their 
disk formation criterion are underestimated by a factor of 2.6.

Figure \ref{fig_compare} shows the evolution of the ratio of circularisation radius 
to the radius of innermost stable circular orbit during the BH+O phase of the 17 
progenitor WR+O binaries, with (\textit{right panel}, Eq.\,\ref{IT96-1}) and without 
(\textit{left panel}, Eq.\,\ref{rd_risco_it96}) taking into consideration the typical 
O star wind velocity, the mass of the O star, and the orbital velocity in the definition 
of accretion radius. Comparing the left and right hand side panels, we see that using 
the appropriate O star wind velocity, the fraction of the BH+O star phase when an 
accretion disk can form greatly decreases. This shows that the X-ray active lifetime 
is very sensitive to the wind velocity considered in the disk formation criterion.

From the modified criterion (Eq.\,\ref{IT96-1}), we see that only 4 out of 17 progenitor WR+O 
binaries are to become wind-fed BH HMXBs for a short period during their 
lifetime as a BH+O star binary. All of them are close binaries with orbital 
periods less than 10 days. The only observed wind-fed BH+O in the Milky Way, 
Cyg X-1, also has an orbital period of around $\sim$5.6 days \citep{orosz2011,Hirsch19}. 
From the modified criterion, we expect to find $\sim$3 wind-fed BH HMXBs 
for the 17 progenitor WR+O binaries instead of 44 as calculated in V20.

In the definition of the accretion radius (Eq.\,\ref{R_BH_van}), 
\citet{iben1996} only account for the wind velocity of the giant star 
companion and not for the relative velocity between the compact object 
and the red giant star. We find that this assumption does not play a 
significant role in most of our BH+O systems as the wind velocity is 
much larger than the orbital velocity (Fig.\,\ref{vw-vorb}). But for 
systems where the wind velocity can be comparable to the orbital velocity, 
the inclusion of the orbital velocity can further reduce the X-ray active 
lifetime. 

\citet{iben1996} also approximated the total mass of the binary 
system to be approximately equal to the mass of the giant star companion 
(see their eq. 65). Their work was primarily aimed at white dwarf/neutron 
star+red giant binary systems and hence this was a reasonable approximation. 
However, that approximation breaks down for BH+O systems. V20 did not correct 
for the mass of the BH in the equation of the orbital velocity. The 
inclusion of the mass of the BH in the expression for orbital velocity 
reduces the predicted X-ray active lifetime of the BH+O models for systems 
where the mass of the BH formed is comparable to the mass of the O star, 
most readily seen from Eq.\,\eqref{correct}. For an equal mass BH+O 
binary, accounting for the mass of the BH introduces a factor of $\sim$1.34 
in the right hand side of Eq.\,\eqref{correct}, which means that the radius 
of the O star has to be larger, all other parameters fixed, for an accretion 
disk to form.

The luminosity of massive O stars can be a finite fraction of its 
Eddington luminosity (see Table \ref{comparison}). Accounting for 
this Eddington factor, defined as the ratio of the luminosity of 
the O star to its Eddington luminosity, in the wind velocity of O 
stars leads to a decrease in the calculated O star wind velocity. 
However, in many of our considered WR+O binaries, the Eddington 
factor of the O star is low ($\leq$ 0.1). Hence, the inclusion of the 
Eddington factor does not have a significant effect on the predicted 
X-ray active lifetime of most of our BH+O models. On the other hand, 
for the few systems which have Eddington factors $\sim$0.3, accounting 
for the Eddington factor increased the X-ray active lifetime, but not 
as significantly as to compensate for the updated O star terminal 
wind velocity.

For the X-ray emission from a BH+O model with an accretion disk to be 
detectable from Earth, V20 assumed a luminosity cut-off of 10$^{35}$ 
erg s$^{-1}$ for all the 17 systems regardless of their individual 
distances from Earth. But most of the 17 WR+O binaries considered 
are not located within 3-4 kpc. In our work, we assume a flux cut-off of 
$\sim$10$^{-11}$ erg s$^{-1}$ cm$^{-2}$, and take the distance 
of each source from Earth into consideration individually. The consideration 
of the individual distances does not affect our results as the calculated 
X-ray flux is above our flux detection threshold for all the models that 
are predicted to have an X-ray active phase (Fig.\,\ref{fiducial_lum_flux}). 

The end of the BH+O phase in V20 is taken to be the point when the 
companion star fills its Roche lobe. This can lead to an over-prediction 
of  $\tau_{\rm Lx}$ for comparatively wide systems where the O 
star can complete hydrogen burning and yet not fill its Roche lobe. 
Since the wind velocity of post MS stars are low as well, some of these 
systems in the post MS phase of the O star can become strong X-ray 
emitters but do not necessarily fall under the class of wind-fed BH HMXBs. 
Hence, V20 may have over-predicted the X-ray active lifetime for some of 
the progenitor WR systems binaries by including the post-MS phase.

A recent population synthesis study by \citet{shao2020} based on rapid 
binary evolution code predicted about 10 - 30 wind-fed BH HMXBs in the 
Milky Way \citep[see also,][]{wiktorowicz2020}. The mass loss rate in 
\citet{vink2001} is adopted, and the accretion rate is evaluated by the 
Bondi-Hoyle-Lyttleton accretion model \citep{bondi1944,Belczynski2008}. 
To evaluate the detectability of their BH+O binary models, they also 
adopt the same threshold for X-ray luminosity at $10^{35}$ erg/s as V20. 
We note that they did not take into account the criterion for the formation 
of an accretion disk. Our work suggests that accretion disks can only exist 
for a limited period of the main sequence lifetime of the O stars, which 
mainly determines the X-ray active lifetime of the BH+O star binaries. 
Therefore, \citet{shao2020} have likely overestimated the number of 
wind-fed BH HMXBs in the Milky Way. 

\section{Discussion}
\label{dnc}
Here, we discuss the uncertainties in the predicted X-ray active lifetimes 
of our BH+O binary models.

\subsection{Specific angular momentum accretion}

The discrepancy between the predicted wind-fed BH HMXB 
populations of V20 and our work shows that the criterion for 
accretion disk formation is sensitive to variations of the 
parameters in the theory. In particular, accounting for a 
larger O star wind velocity changes the prediction of V20 
drastically. More factors such as the accretion efficiency, and the 
approximation of the specific angular momentum carried 
by the accreted matter introduce further uncertainties 
for the computation of the X-Ray bright lifetime of the 
BH+O star binaries (Sect.\,\ref{results}). \citet{livio1986} 
and \citet{soker1986} studied accretion onto compact objects 
using detailed hydrodynamic simulations and found that while 
the mass accretion rate is similar to that predicted by the 
Bondi-Hoyle theory, the amount of specific angular momentum 
accreted was only a few percent of that predicted by the 
analytical approximation obtained from the Bondi-Hoyle theory. 
A similar conclusion was drawn by \citet{ruffert1999}. 

Owing to these calculations, \citet{elmellahthesis} adopted 
the analytical expression for specific angular momentum from 
\citet[][eq. 7]{shapiro1976} but introduced an efficiency factor 
of 1/3 to account for the reduced specific angular momentum 
accretion found in the detailed numerical hydrodynamic studies. 
We captured this uncertainty and studied its effects through our 
efficiency parameter $\eta$. Therefore, we need to compare the 
analytical approximations to the specific angular momentum 
carried by the accreted matter used in V20 and our work, to 
detailed 3D numerical hydrodynamic simulations to assess 
the reliability of the approximations.

\begin{figure}
\centering
\includegraphics[width=\hsize]{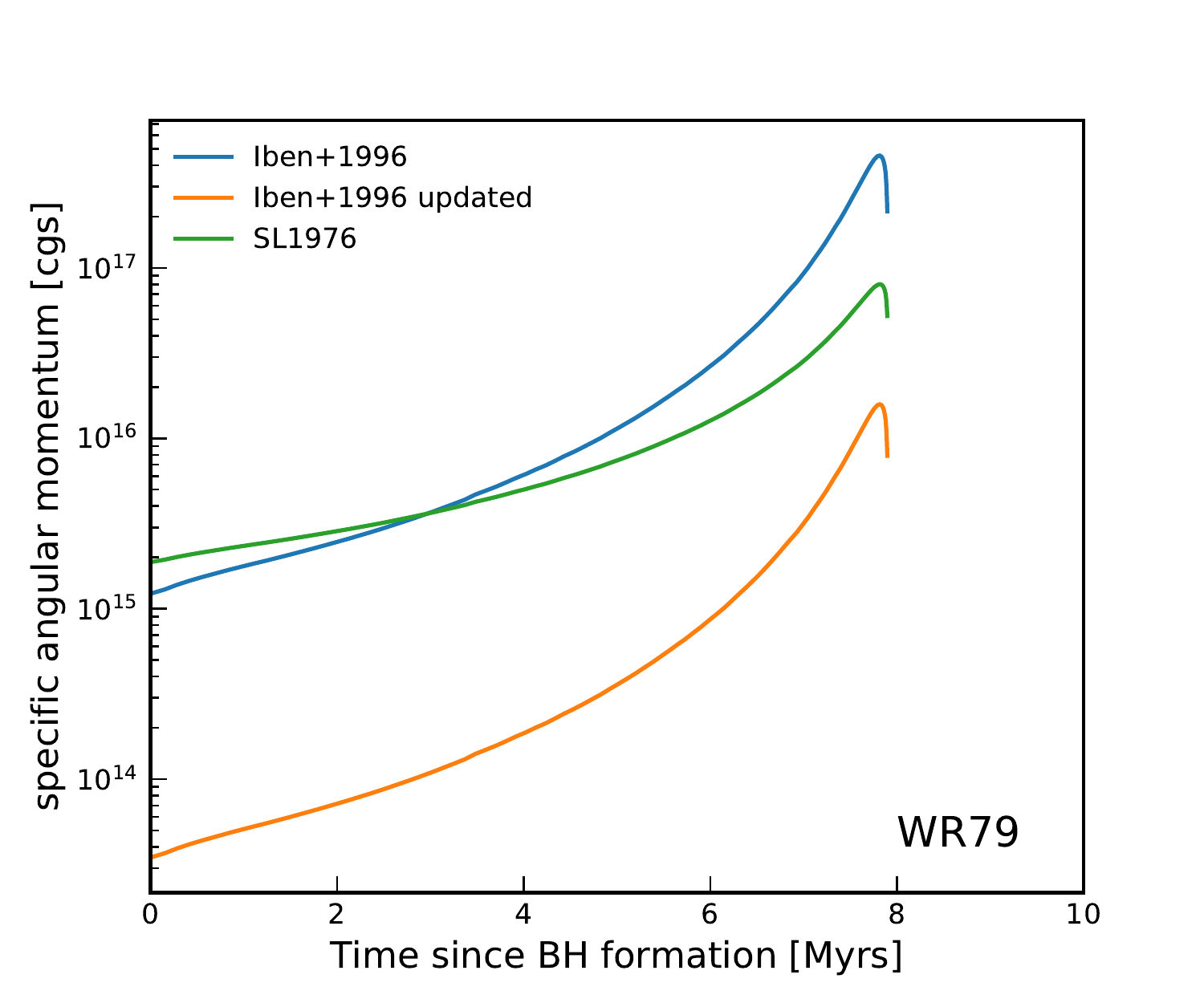} 
\caption{Comparison of the specific angular momentum of the accreted wind matter used in V20 (Eq.\,\ref{specific_j_van}, Iben+1996), to the analytical value derived by \citet{shapiro1976} (SL1976), for the stellar and binary parameters of WR79, as a function of the time since black hole formation. We also show the decrease in the amount of specific angular momentum carried by the wind when the typical wind velocity for O type stars is introduced in Eq.\,\eqref{specific_j_van}, denoted by `Iben+1996 updated'.}
\label{j_acc_compare}
\end{figure}

As a preliminary exercise, in Fig.\,\ref{j_acc_compare}, we compare 
the expression for specific angular momentum carried by the wind 
matter used in V20 \citep[eq. 62 of][]{iben1996} to the analytical 
form derived by \citet[][]{shapiro1976} that is assumed in our work. 
We see that there are significant differences between the two definitions, 
and to the third, which includes the typical O star wind velocity 
in expression for specific angular momentum carried by the wind matter 
given by \citet{iben1996}.

Due to the line-deshadowing instability and sub-photospheric 
turbulence, stellar winds from hot stars are prone to form 
overdense regions called “clumps” \citep{Owocki1984,Owocki1988,Feldmeier1995,Grassitelli2015}. 
These clumps produce stochastic variations in the instantaneous 
amount of specific angular momentum of the accreted material. 
These variations take place on time scales of the order of 
hundreds to thousands of seconds, much shorter than the evolutionary 
time scales \citep{Grinberg2017,ElMellah2020}. For clump 
sizes derived from first principles \citep{Sundqvist2018}, 
clumps are small compared to the accretion radius when they 
reach the orbital separation \citep{elmellah2018}. As a 
consequence, they induce a limited peak-to-peak variability. 
However, when the wind is sufficiently fast, the net amount 
of angular momentum provided to the flow is so small (and so 
is the accretion radius) that the serendipitous capture of 
clumps becomes relatively more important and can produce a 
transient accretion disk. However, in Cygnus X-1, the 
wind-captured disk is permanent and so far, the only wind-fed 
HMXB where a transient wind-captured disk was observed is 
Vela X-1 \citep{Liao2020}. In the latter case, the disk 
formation is believed to be associated to variations at the 
periastron induced by the slightly eccentric orbital motion, 
rather than to clump capture \citep{Kretschmar2021}. 
Therefore, including wind clumping is not expected to 
significantly modify the results obtained in this paper.

\subsection{Properties of the WR star companion}

In many of the investigated WR+O star binaries, in particular 
in the shorter-period ones, the WR star likely formed via 
Roche-lobe overflow from its progenitor O star 
\citep[e.g.][]{vanbeveren1998b,Wellstein1999}. The companion 
O star may thus accrete mass from the WR star progenitor, 
which could lead to properties which are different from those 
of single O stars. Important properties in this respect are the 
helium abundance and spin of the mass gaining O star.

An enhancement of the surface helium mass fraction of the mass 
gainer of a few percent is predicted from conservative \citep{Wellstein1999} 
as well as non-conservative \citep{Petrovic2005a,langer2020} 
massive binary evolution models. This enrichment leads to a 
slight overluminosity of the mass gainer \citep{langer1992}, 
which may affect the stellar wind properties. However, 
quantitatively, this effect is not expected to exceed the 
uncertainty in the average wind properties of O stars \citep{vink2021}.

Independent of the mass transfer efficiency, the angular 
momentum gain of the accretor during the mass transfer is 
expected to spin-up the mass gainer significantly 
\citep{packet1981,Petrovic2005b,langer2020}. The observed 
population of Be/X-ray binaries \citep{Reig2011} signifies 
that this spin-up may achieve near-critical rotation, with 
strong consequences for the mass outflow from the spun-up 
star, and the mass accretion onto the compact companion. 
The Galactic and LMC WR+O binaries indeed also contain 
rapidly rotating O stars \citep{vanbeveren2018,shara2020}. 
However, while faster than average O stars, the analysed 
WR companions rotate on average with less than 50\% of their 
critical rotational velocity, implying that the centrifugal 
force remains below 25\% of the surface gravity at the stellar 
equator. Whereas this may lead to a slight wind anisotropy, 
a disk-like outflow is not expected in this case. 

In our analysis above, we adopted a wind velocity of the O 
star companions as expected for single stars. However, in 
Be/X-ray binaries \citep{waters1988} as well as in supergiant 
X-ray binaries \citep{Manousakis2012}, abnormally slow stellar 
winds are observed. While in the first case, this may relate 
to the stars' extreme rotation, a reduced wind acceleration 
due to the X-ray irradiation of the stellar atmosphere is 
thought to be responsible in the latter case \citep[see also,][]{Vilhu2021}. 
For the conclusions we draw above, the consequences would be 
small, since none of the two effects is expected in the 
majority of the investigated binaries. For the few cases 
where disk formation and significant X-ray emission is 
predicted, a slower wind would, however, lead to an increased 
accretion rate and a higher X-ray luminosity.

\subsection{Other uncertainties}

The lifetime of WR+O star binary phase is taken to be 
constant for all the different considered WR+O binaries, 
whereas it actually depends on the core helium burning 
lifetime of the WR star, which in turn depends on the 
individual masses of the WR star. However, we do not 
expect this simplifying assumption to affect our predicted 
number of wind-fed BH HMXBs significantly. 
The mass of the WR stars at the end of core helium depletion 
is also uncertain due to the uncertainty in the mass 
loss rate during the WR phase \citep{neijssel2021}. In both works, 
ours and V20, it is assumed that the properties of the WR stars do not 
change after core helium depletion. However, it has been 
shown recently \citep{laplace2020} that WR stars that have 
an outer hydrogen envelope may expand post helium depletion 
and the binary can undergo another mass transfer phase before 
core collapse \citep[see also][]{laplace2021}. 
Therefore, the formation of wind-fed BH HMXBs needs further 
investigation, both using detailed binary evolution models 
that calculate the binary evolution up to the core collapse 
of the WR star as well as into accurate modelling of the 
physics of accretion onto BHs.

We have shown (Fig.\,\ref{mdot_acc}) that the predicted 
mass accretion rates calculated for the anticipated 
BH+O binaries are much lower than the Eddington mass 
accretion rates. Hence, we don't consider super-Eddington 
accretion to be relevant to our work. Due to the same reason, 
the X-ray emission should be isotropic and we do not need 
to consider the case of beaming \citep{king2008}. 
LOBSTER eye telescopes can reach a flux cut-off of 
$10^{-12}$ erg s$^{-1}$ cm$^{-2}$ \citep{Priedhorsky1996,Hudec2007}. 
The recently launched eROSITA X-ray telescope is stated to have a 
flux detection threshold of $\sim$10$^{-14}$ erg s$^{-1}$ cm$^{-2}$ 
\citep{Merloni2012} in the average all-sky survey mode. 
However, changing the flux limit to these lower values does 
not change our predicted X-ray active lifetime as once a wind-captured 
disk can form around the BH, the expected to X-ray flux at Earth 
is higher than 10$^{-11}$ erg s$^{-1}$ cm$^{-2}$ (Fig.\,\ref{fiducial_lum_flux}). 
On the other hand, if there is significant extinction at X-ray 
wavelengths in the Galactic plane, our predicted X-ray active 
lifetime of the BH+O binary models can get reduced.

The predicted X-ray luminosity of an accreting BH in our models 
is a few percent of the Eddington luminosity. In such a case, the X-ray spectrum 
can switch from a soft state to a hard state and the X-ray emission 
becomes radiatively inefficient \citep{Yuan2014}. This is well-described 
by a distended and tenuous advection-dominated accretion flow 
\citep[ADAF, see][]{Narayan1994,Narayan1995a,Narayan1995b}. In 
this accretion regime, the bulk of the accretion energy is carried 
by the accreting gas in the form of thermal energy, that can vanish 
through the event horizon of the BH. Hence, the black holes in the 
ADAF regime can be fainter by a factor of $\sim$100-1000 \citep{Narayan2008}. 
In such a case, we do not expect the binary to be detectable in X-rays. 
Inclusion of this effect can only reduce our prediction of the X-ray 
active lifetime during the BH+O phase.

\section{Conclusion}
\label{conclusion}
WR+O binaries are expected to be progenitors of BH+O 
binaries. V20 investigated 17 galactic WR+O star binaries 
and predicted that there should be more than 200 wind-fed BH HMXBs in the 
Milky Way while only one has been observed. They concluded that 
BHs receive much higher natal kick velocities or WR stars 
explode with supernova explosions to form neutron stars, 
which lead to a break-up of the binary systems. 

We apply a similar methodology as in V20 with an improved analytical 
criterion to study the formation of accretion disks around BHs in 
BH+O binaries and the detectability of X-ray emission from such systems. 
We also investigate the effect of uncertain physics 
parameters, such as the $\beta$ value in the O star wind velocity 
law, the efficiency of angular momentum accretion ($\eta$) and 
the spin of the BH ($\gamma_{\pm}$), on the predicted number of 
wind-fed BH HMXBs. We find that this calculated number 
is sensitive to plausible variations of the assumed parameters.

For our fiducial parameter set ($\beta$,\,$\eta$,\,$\gamma$) = (1,\,1/3,\,1) 
(see Sect.\,\ref{results},and Fig.\,\ref{plot_result}), we predict only $\sim$2-3 
wind-fed BH HMXBs based on the 17 progenitor WR+O systems. 
While we still over-predict the number of wind-fed BH HMXBs, 
accounting for the theoretical and observational biases in the population of WR+O binaries 
(see Sect.\,\ref{sample}) suggest that we should expect $\sim$2-3 
wind-fed BH HMXBs in the entire Milky Way. We remind the 
reader that only 1 wind-fed BH+O X-ray binary has been observed (Cyg X-1). 

We then revisited the derivation of the accretion disk formation 
criterion used by V20 and found that, in particular, the assumed 
O star wind velocity was underestimated. Accounting for the 
appropriate O star wind velocity \citep{vink2001}, we find 
most of BH+O binary models will have negligible X-ray bright 
lifetimes due to the absence of an accretion disk around the BH 
(see Fig.\,\ref{fig_compare}). As such, any conclusion drawn from 
the seemingly discrepant number of observed WR+O binaries and 
wind-fed BH HMXBs has to be re-evaluated. 

Our analysis shows further that a high BH spin parameter can lead 
to significantly longer and brighter X-ray phases in wind-accreting 
BH+O binaries. The corresponding bias in detecting such binaries 
with rapidly spinning BHs may help to alleviate the tension between 
the rather low BH spin values generally predicted from binary stellar 
evolution models \citep{qin2018}, and the high BH spin values 
observationally deduced from BH+O binaries in the Local Group \citep{qin2019}. 

We conclude that high BH formation kicks are not 
necessary to understand the number discrepancy between the 
populations of observed WR+O binaries and wind-fed BH HMXBs 
in the Milky Way. 
With our current understanding of O star wind velocities, 
we have shown that possibly the vast majority of Galactic 
BH+O star binaries may not form BH accretion disks and hence remain 
undetected in X-Ray surveys. Recent studies have shown that the 
GAIA satellite offers an excellent opportunity to observe X-ray 
such quiet BH+O binaries via periodic astrometric variations \citep{Breivik2017,Mashian2017,Yalinewich2018,Yamaguchi2018,Andrews2019}. 
Furthermore, BH+O binaries can also be detected from photometric 
variability of the O star induced by the BH companion \citep{zucker2007,masuda2019}, 
or spectroscopically via the periodic shift in radial velocity 
of the O star.

\begin{acknowledgements}

We thank Pablo Marchant and Krzysztof Belczynski for meaningful 
discussions, and Dany Vanbeveren for helpful comments on an
earlier version of this manuscript. I.E.M. has received funding 
from the European Research Council (ERC) under the European Union’s 
Horizon 2020 research and innovation programme (Spawn ERC, grant 
agreement No 863412). This research has made use of NASA’s 
Astrophysics Data System. 

\end{acknowledgements}

\bibliographystyle{aa}
\bibliography{p4}

\begin{thebibliography}{134}
\expandafter\ifx\csname natexlab\endcsname\relax\def\natexlab#1{#1}\fi

\bibitem[{{Abbott} {et~al.}(2016){Abbott}, {Abbott}, {Abbott}, {Abernathy},
  {Acernese}, {Ackley}, {Adams}, {Adams}, {Addesso}, {Adhikari}, {Adya},
  {Affeldt}, {Agathos}, {Agatsuma}, {Aggarwal}, {Aguiar}, {Aiello}, {Ain},
  {Ajith}, {Allen}, {Allocca}, {Altin}, {Anderson}, {Anderson}, {Arai},
  {Araya}, {Arceneaux}, {Areeda}, {Arnaud}, {Arun}, {Ascenzi}, {Ashton}, {Ast},
  {Aston}, {Astone}, {Aufmuth}, {Aulbert}, {Babak}, {Bacon}, {Bader}, {Baker},
  {Baldaccini}, {Ballardin}, {Ballmer}, {Barayoga}, {Barclay}, {Barish},
  {Barker}, {Barone}, {Barr}, {Barsotti}, {Barsuglia}, {Barta}, {Bartlett},
  {Bartos}, {Bassiri}, {Basti}, {Batch}, {Baune}, {Bavigadda}, {Bazzan},
  {Bejger}, {Bell}, {Berger}, {Bergmann}, {Berry}, {Bersanetti}, {Bertolini},
  {Betzwieser}, {Bhagwat}, {Bhandare}, {Bilenko}, {Billingsley}, {Birch},
  {Birney}, {Birnholtz}, {Biscans}, {Bisht}, {Bitossi}, {Biwer}, {Bizouard},
  {Blackburn}, {Blair}, {Blair}, {Blair}, {Bloemen}, {Bock}, {Boer}, {Bogaert},
  {Bogan}, {Bohe}, {Bond}, {Bondu}, {Bonnand}, {Boom}, {Bork}, {Boschi},
  {Bose}, {Bouffanais}, {Bozzi}, {Bradaschia}, {Brady}, {Braginsky},
  {Branchesi}, {Brau}, {Briant}, {Brillet}, {Brinkmann}, {Brisson}, {Brockill},
  {Broida}, {Brooks}, {Brown}, {Brown}, {Brown}, {Brunett}, {Buchanan},
  {Buikema}, {Bulik}, {Bulten}, {Buonanno}, {Buskulic}, {Buy}, {Byer},
  {Cabero}, {Cadonati}, {Cagnoli}, {Cahillane}, {Calder{\'o}n Bustillo},
  {Callister}, {Calloni}, {Camp}, {Cannon}, {Cao}, {Capano}, {Capocasa},
  {Carbognani}, {Caride}, {Casanueva Diaz}, {Casentini}, {Caudill},
  {Cavagli{\`a}}, {Cavalier}, {Cavalieri}, {Cella}, {Cepeda}, {Cerboni
  Baiardi}, {Cerretani}, {Cesarini}, {Chamberlin}, {Chan}, {Chao}, {Charlton},
  {Chassande-Mottin}, {Cheeseboro}, {Chen}, {Chen}, {Cheng}, {Chincarini},
  {Chiummo}, {Cho}, {Cho}, {Chow}, {Christensen}, {Chu}, {Chua}, {Chung},
  {Ciani}, {Clara}, {Clark}, {Cleva}, {Coccia}, {Cohadon}, {Colla}, {Collette},
  {Cominsky}, {Constancio}, {Conte}, {Conti}, {Cook}, {Corbitt}, {Cornish},
  {Corsi}, {Cortese}, {Costa}, {Coughlin}, {Coughlin}, {Coulon}, {Countryman},
  {Couvares}, {Cowan}, {Coward}, {Cowart}, {Coyne}, {Coyne}, {Craig},
  {Creighton}, {Cripe}, {Crowder}, {Cumming}, {Cunningham}, {Cuoco}, {Dal
  Canton}, {Danilishin}, {D'Antonio}, {Danzmann}, {Darman}, {Dasgupta}, {Da
  Silva Costa}, {Dattilo}, {Dave}, {Davier}, {Davies}, {Daw}, {Day}, {De},
  {DeBra}, {Debreczeni}, {Degallaix}, {De Laurentis}, {Del{\'e}glise}, {Del
  Pozzo}, {Denker}, {Dent}, {Dergachev}, {De Rosa}, {DeRosa}, {DeSalvo},
  {Devine}, {Dhurand har}, {D{\'\i}az}, {Di Fiore}, {Di Giovanni}, {Di
  Girolamo}, {Di Lieto}, {Di Pace}, {Di Palma}, {Di Virgilio}, {Dolique},
  {Donovan}, {Dooley}, {Doravari}, {Douglas}, {Downes}, {Drago}, {Drever},
  {Driggers}, {Ducrot}, {Dwyer}, {Edo}, {Edwards}, {Effler}, {Eggenstein},
  {Ehrens}, {Eichholz}, {Eikenberry}, {Engels}, {Essick}, {Etzel}, {Evans},
  {Evans}, {Everett}, {Factourovich}, {Fafone}, {Fair}, {Fairhurst}, {Fan},
  {Fang}, {Farinon}, {Farr}, {Farr}, {Favata}, {Fays}, {Fehrmann}, {Fejer},
  {Fenyvesi}, {Ferrante}, {Ferreira}, {Ferrini}, {Fidecaro}, {Fiori},
  {Fiorucci}, {Fisher}, {Flaminio}, {Fletcher}, {Fong}, {Fournier}, {Frasca},
  {Frasconi}, {Frei}, {Freise}, {Frey}, {Frey}, {Fritschel}, {Frolov}, {Fulda},
  {Fyffe}, {Gabbard}, {Gaebel}, {Gair}, {Gammaitoni}, {Gaonkar}, {Garufi},
  {Gaur}, {Gehrels}, {Gemme}, {Geng}, {Genin}, {Gennai}, {George}, {Gergely},
  {Germain}, {Ghosh}, {Ghosh}, {Ghosh}, {Giaime}, {Giardina}, {Giazotto},
  {Gill}, {Glaefke}, {Goetz}, {Goetz}, {Gondan}, {Gonz{\'a}lez}, {Gonzalez
  Castro}, {Gopakumar}, {Gordon}, {Gorodetsky}, {Gossan}, {Gosselin}, {Gouaty},
  {Grado}, {Graef}, {Graff}, {Granata}, {Grant}, {Gras}, {Gray}, {Greco},
  {Green}, {Groot}, {Grote}, {Grunewald}, {Guidi}, {Guo}, {Gupta}, {Gupta},
  {Gushwa}, {Gustafson}, {Gustafson}, {Hacker}, {Hall}, {Hall}, {Hamilton},
  {Hammond}, {Haney}, {Hanke}, {Hanks}, {Hanna}, {Hannam}, {Hanson},
  {Hardwick}, {Harms}, {Harry}, {Harry}, {Hart}, {Hartman}, {Haster},
  {Haughian}, {Healy}, {Heidmann}, {Heintze}, {Heitmann}, {Hello}, {Hemming},
  {Hendry}, {Heng}, {Hennig}, {Henry}, {Heptonstall}, {Heurs}, {Hild}, {Hoak},
  {Hofman}, {Holt}, {Holz}, {Hopkins}, {Hough}, {Houston}, {Howell}, {Hu},
  {Huang}, {Huerta}, {Huet}, {Hughey}, {Husa}, {Huttner}, {Huynh-Dinh},
  {Indik}, {Ingram}, {Inta}, {Isa}, {Isac}, {Isi}, {Isogai}, {Iyer}, {Izumi},
  {Jacqmin}, {Jang}, {Jani}, {Jaranowski}, {Jawahar}, {Jian},
  {Jim{\'e}nez-Forteza}, {Johnson}, {Johnson-McDaniel}, {Jones}, {Jones},
  {Jonker}, {Ju}, {K}, {Kalaghatgi}, {Kalogera}, {Kandhasamy}, {Kang},
  {Kanner}, {Kapadia}, {Karki}, {Karvinen}, {Kasprzack}, {Katsavounidis},
  {Katzman}, {Kaufer}, {Kaur}, {Kawabe}, {K{\'e}f{\'e}lian}, {Kehl}, {Keitel},
  {Kelley}, {Kells}, {Kennedy}, {Key}, {Khalili}, {Khan}, {Khan}, {Khan},
  {Khazanov}, {Kijbunchoo}, {Kim}, {Kim}, {Kim}, {Kim}, {Kim}, {Kim}, {Kim},
  {Kimbrell}, {King}, {King}, {Kissel}, {Klein}, {Kleybolte}, {Klimenko},
  {Koehlenbeck}, {Koley}, {Kondrashov}, {Kontos}, {Korobko}, {Korth},
  {Kowalska}, {Kozak}, {Kringel}, {Krishnan}, {Kr{\'o}lak}, {Krueger}, {Kuehn},
  {Kumar}, {Kumar}, {Kuo}, {Kutynia}, {Lackey}, {Land ry}, {Lange}, {Lantz},
  {Lasky}, {Laxen}, {Lazzarini}, {Lazzaro}, {Leaci}, {Leavey}, {Lebigot},
  {Lee}, {Lee}, {Lee}, {Lee}, {Lenon}, {Leonardi}, {Leong}, {Leroy},
  {Letendre}, {Levin}, {Lewis}, {Li}, {Libson}, {Littenberg}, {Lockerbie},
  {Lombardi}, {London}, {Lord}, {Lorenzini}, {Loriette}, {Lormand}, {Losurdo},
  {Lough}, {Lousto}, {L{\"u}ck}, {Lundgren}, {Lynch}, {Ma}, {Machenschalk},
  {MacInnis}, {Macleod}, {Maga{\~n}a-Sandoval}, {Maga{\~n}a Zertuche}, {Magee},
  {Majorana}, {Maksimovic}, {Malvezzi}, {Man}, {Mandel}, {Mandic}, {Mangano},
  {Mansell}, {Manske}, {Mantovani}, {Marchesoni}, {Marion}, {M{\'a}rka},
  {M{\'a}rka}, {Markosyan}, {Maros}, {Martelli}, {Martellini}, {Martin},
  {Martynov}, {Marx}, {Mason}, {Masserot}, {Massinger}, {Masso-Reid},
  {Mastrogiovanni}, {Matichard}, {Matone}, {Mavalvala}, {Mazumder}, {McCarthy},
  {McClelland}, {McCormick}, {McGuire}, {McIntyre}, {McIver}, {McManus},
  {McRae}, {McWilliams}, {Meacher}, {Meadors}, {Meidam}, {Melatos}, {Mendell},
  {Mercer}, {Merilh}, {Merzougui}, {Meshkov}, {Messenger}, {Messick},
  {Metzdorff}, {Meyers}, {Mezzani}, {Miao}, {Michel}, {Middleton}, {Mikhailov},
  {Milano}, {Miller}, {Miller}, {Miller}, {Miller}, {Millhouse}, {Minenkov},
  {Ming}, {Mirshekari}, {Mishra}, {Mitra}, {Mitrofanov}, {Mitselmakher},
  {Mittleman}, {Moggi}, {Mohan}, {Mohapatra}, {Montani}, {Moore}, {Moore},
  {Moraru}, {Moreno}, {Morriss}, {Mossavi}, {Mours}, {Mow-Lowry}, {Mueller},
  {Muir}, {Mukherjee}, {Mukherjee}, {Mukherjee}, {Mukund}, {Mullavey}, {Munch},
  {Murphy}, {Murray}, {Mytidis}, {Nardecchia}, {Naticchioni}, {Nayak},
  {Nedkova}, {Nelemans}, {Nelson}, {Neri}, {Neunzert}, {Newton}, {Nguyen},
  {Nielsen}, {Nissanke}, {Nitz}, {Nocera}, {Nolting}, {Normandin}, {Nuttall},
  {Oberling}, {Ochsner}, {O'Dell}, {Oelker}, {Ogin}, {Oh}, {Oh}, {Ohme},
  {Oliver}, {Oppermann}, {Oram}, {O'Reilly}, {O'Shaughnessy}, {Ottaway},
  {Overmier}, {Owen}, {Pai}, {Pai}, {Palamos}, {Palashov}, {Palomba},
  {Pal-Singh}, {Pan}, {Pan}, {Pankow}, {Pannarale}, {Pant}, {Paoletti},
  {Paoli}, {Papa}, {Paris}, {Parker}, {Pascucci}, {Pasqualetti}, {Passaquieti},
  {Passuello}, {Patricelli}, {Patrick}, {Pearlstone}, {Pedraza}, {Pedurand},
  {Pekowsky}, {Pele}, {Penn}, {Perreca}, {Perri}, {Pfeiffer}, {Phelps},
  {Piccinni}, {Pichot}, {Piergiovanni}, {Pierro}, {Pillant}, {Pinard}, {Pinto},
  {Pitkin}, {Poe}, {Poggiani}, {Popolizio}, {Porter}, {Post}, {Powell},
  {Prasad}, {Predoi}, {Prestegard}, {Price}, {Prijatelj}, {Principe},
  {Privitera}, {Prix}, {Prodi}, {Prokhorov}, {Puncken}, {Punturo}, {Puppo},
  {P{\"u}rrer}, {Qi}, {Qin}, {Qiu}, {Quetschke}, {Quintero}, {Quitzow-James},
  {Raab}, {Rabeling}, {Radkins}, {Raffai}, {Raja}, {Rajan}, {Rakhmanov},
  {Rapagnani}, {Raymond}, {Razzano}, {Re}, {Read}, {Reed}, {Regimbau}, {Rei},
  {Reid}, {Reitze}, {Rew}, {Reyes}, {Ricci}, {Riles}, {Rizzo}, {Robertson},
  {Robie}, {Robinet}, {Rocchi}, {Rolland}, {Rollins}, {Roma}, {Romano},
  {Romano}, {Romanov}, {Romie}, {Rosi{\'n}ska}, {Rowan}, {R{\"u}diger},
  {Ruggi}, {Ryan}, {Sachdev}, {Sadecki}, {Sadeghian}, {Sakellariadou},
  {Salconi}, {Saleem}, {Salemi}, {Samajdar}, {Sammut}, {Sanchez}, {Sandberg},
  {Sandeen}, {Sand ers}, {Sassolas}, {Sathyaprakash}, {Saulson}, {Sauter},
  {Savage}, {Sawadsky}, {Schale}, {Schilling}, {Schmidt}, {Schmidt},
  {Schnabel}, {Schofield}, {Sch{\"o}nbeck}, {Schreiber}, {Schuette}, {Schutz},
  {Scott}, {Scott}, {Sellers}, {Sengupta}, {Sentenac}, {Sequino}, {Sergeev},
  {Setyawati}, {Shaddock}, {Shaffer}, {Shahriar}, {Shaltev}, {Shapiro},
  {Shawhan}, {Sheperd}, {Shoemaker}, {Shoemaker}, {Siellez}, {Siemens},
  {Sieniawska}, {Sigg}, {Silva}, {Singer}, {Singer}, {Singh}, {Singh},
  {Singhal}, {Sintes}, {Slagmolen}, {Smith}, {Smith}, {Smith}, {Son}, {Sorazu},
  {Sorrentino}, {Souradeep}, {Srivastava}, {Staley}, {Steinke}, {Steinlechner},
  {Steinlechner}, {Steinmeyer}, {Stephens}, {Stevenson}, {Stone}, {Strain},
  {Straniero}, {Stratta}, {Strauss}, {Strigin}, {Sturani}, {Stuver},
  {Summerscales}, {Sun}, {Sunil}, {Sutton}, {Swinkels}, {Szczepa{\'n}czyk},
  {Tacca}, {Talukder}, {Tanner}, {T{\'a}pai}, {Tarabrin}, {Taracchini},
  {Taylor}, {Theeg}, {Thirugnanasamband am}, {Thomas}, {Thomas}, {Thomas},
  {Thorne}, {Thrane}, {Tiwari}, {Tiwari}, {Tokmakov}, {Toland}, {Tomlinson},
  {Tonelli}, {Tornasi}, {Torres}, {Torrie}, {T{\"o}yr{\"a}}, {Travasso},
  {Traylor}, {Trifir{\`o}}, {Tringali}, {Trozzo}, {Tse}, {Turconi},
  {Tuyenbayev}, {Ugolini}, {Unnikrishnan}, {Urban}, {Usman}, {Vahlbruch},
  {Vajente}, {Valdes}, {Vallisneri}, {van Bakel}, {van Beuzekom}, {van den
  Brand}, {Van Den Broeck}, {Vand er-Hyde}, {van der Schaaf}, {van Heijningen},
  {van Veggel}, {Vardaro}, {Vass}, {Vas{\'u}th}, {Vaulin}, {Vecchio},
  {Vedovato}, {Veitch}, {Veitch}, {Venkateswara}, {Verkindt}, {Vetrano},
  {Vicer{\'e}}, {Vinciguerra}, {Vine}, {Vinet}, {Vitale}, {Vo}, {Vocca},
  {Vorvick}, {Voss}, {Vousden}, {Vyatchanin}, {Wade}, {Wade}, {Wade}, {Walker},
  {Wallace}, {Walsh}, {Wang}, {Wang}, {Wang}, {Wang}, {Wang}, {Ward}, {Warner},
  {Was}, {Weaver}, {Wei}, {Weinert}, {Weinstein}, {Weiss}, {Wen}, {We{\ss}els},
  {Westphal}, {Wette}, {Whelan}, {Whitcomb}, {Whiting}, {Williams},
  {Williamson}, {Willis}, {Willke}, {Wimmer}, {Winkler}, {Wipf}, {Wittel},
  {Woan}, {Woehler}, {Worden}, {Wright}, {Wu}, {Wu}, {Yablon}, {Yam},
  {Yamamoto}, {Yancey}, {Yu}, {Yvert}, {Zadro{\.Z}ny}, {Zangrando}, {Zanolin},
  {Zendri}, {Zevin}, {Zhang}, {Zhang}, {Zhang}, {Zhao}, {Zhou}, {Zhou}, {Zhu},
  {Zucker}, {Zuraw}, {Zweizig}, {LIGO Scientific Collaboration}, \& {Virgo
  Collaboration}}]{abbott2016}
{Abbott}, B.~P., {Abbott}, R., {Abbott}, T.~D., {et~al.} 2016, Physical Review
  X, 6, 041015

\bibitem[{{Abbott} {et~al.}(2019){Abbott}, {Abbott}, {Abbott}, {Abraham},
  {Acernese}, {Ackley}, {Adams}, {Adhikari}, {Adya}, {Affeldt}, {Agathos},
  {Agatsuma}, {Aggarwal}, {Aguiar}, {Aiello}, {Ain}, {Ajith}, {Allen},
  {Allocca}, {Aloy}, \& {Virgo Collaboration}}]{abbott2019}
{Abbott}, B.~P., {Abbott}, R., {Abbott}, T.~D., {et~al.} 2019, Physical Review
  X, 9, 031040

\bibitem[{{Andrews} {et~al.}(2019){Andrews}, {Breivik}, \&
  {Chatterjee}}]{Andrews2019}
{Andrews}, J.~J., {Breivik}, K., \& {Chatterjee}, S. 2019, \apj, 886, 68

\bibitem[{{Bachetti} {et~al.}(2014){Bachetti}, {Harrison}, {Walton},
  {Grefenstette}, {Chakrabarty}, {F{\"u}rst}, {Barret}, {Beloborodov}, {Boggs},
  {Christensen}, {Craig}, {Fabian}, {Hailey}, {Hornschemeier}, {Kaspi},
  {Kulkarni}, {Maccarone}, {Miller}, {Rana}, {Stern}, {Tendulkar}, {Tomsick},
  {Webb}, \& {Zhang}}]{Bachetti2014}
{Bachetti}, M., {Harrison}, F.~A., {Walton}, D.~J., {et~al.} 2014, \nat, 514,
  202

\bibitem[{{Balbus} \& {Hawley}(1991)}]{Balbus1991}
{Balbus}, S.~A. \& {Hawley}, J.~F. 1991, \apj, 376, 214

\bibitem[{{Belczynski} {et~al.}(2014){Belczynski}, {Buonanno}, {Cantiello},
  {Fryer}, {Holz}, {Mand el}, {Miller}, \& {Walczak}}]{belczynski2014}
{Belczynski}, K., {Buonanno}, A., {Cantiello}, M., {et~al.} 2014, \apj, 789,
  120

\bibitem[{Belczynski {et~al.}(2008)Belczynski, Kalogera, Rasio, Taam, Zezas,
  Bulik, Maccarone, \& Ivanova}]{Belczynski2008}
Belczynski, K., Kalogera, V., Rasio, F.~A., {et~al.} 2008, The Astrophysical
  Journal Supplement Series, 174, 223

\bibitem[{{Belczynski} {et~al.}(2020){Belczynski}, {Klencki}, {Fields},
  {Olejak}, {Berti}, {Meynet}, {Fryer}, {Holz}, {O'Shaughnessy}, {Brown},
  {Bulik}, {Leung}, {Nomoto}, {Madau}, {Hirschi}, {Kaiser}, {Jones}, {Mondal},
  {Chruslinska}, {Drozda}, {Gerosa}, {Doctor}, {Giersz}, {Ekstrom}, {Georgy},
  {Askar}, {Baibhav}, {Wysocki}, {Natan}, {Farr}, {Wiktorowicz}, {Coleman
  Miller}, {Farr}, \& {Lasota}}]{belczynski2020}
{Belczynski}, K., {Klencki}, J., {Fields}, C.~E., {et~al.} 2020, \aap, 636,
  A104

\bibitem[{{Belczynski} {et~al.}(2016){Belczynski}, {Repetto}, {Holz},
  {O'Shaughnessy}, {Bulik}, {Berti}, {Fryer}, \& {Dominik}}]{belczynski2016}
{Belczynski}, K., {Repetto}, S., {Holz}, D.~E., {et~al.} 2016, \apj, 819, 108

\bibitem[{{Belczynski} {et~al.}(2012){Belczynski}, {Wiktorowicz}, {Fryer},
  {Holz}, \& {Kalogera}}]{belczynski2012}
{Belczynski}, K., {Wiktorowicz}, G., {Fryer}, C.~L., {Holz}, D.~E., \&
  {Kalogera}, V. 2012, \apj, 757, 91

\bibitem[{{Bondi}(1952)}]{Bondi1952}
{Bondi}, H. 1952, \mnras, 112, 195

\bibitem[{{Bondi} \& {Hoyle}(1944)}]{bondi1944}
{Bondi}, H. \& {Hoyle}, F. 1944, \mnras, 104, 273

\bibitem[{{Bradt} {et~al.}(1991){Bradt}, {Swank}, \& {Rothschild}}]{bradt1991}
{Bradt}, H.~V., {Swank}, J.~H., \& {Rothschild}, R.~E. 1991, Advances in Space
  Research, 11, 243

\bibitem[{{Brandt} {et~al.}(1995){Brandt}, {Podsiadlowski}, \&
  {Sigurdsson}}]{brandt1995}
{Brandt}, W.~N., {Podsiadlowski}, P., \& {Sigurdsson}, S. 1995, \mnras, 277,
  L35

\bibitem[{{Braun} \& {Langer}(1995)}]{braun1995}
{Braun}, H. \& {Langer}, N. 1995, \aap, 297, 483

\bibitem[{{Breivik} {et~al.}(2017){Breivik}, {Chatterjee}, \&
  {Larson}}]{Breivik2017}
{Breivik}, K., {Chatterjee}, S., \& {Larson}, S.~L. 2017, \apjl, 850, L13

\bibitem[{{Crowther}(2015)}]{crowther2015}
{Crowther}, P.~A. 2015, in Wolf-Rayet Stars, ed. W.-R. {Hamann}, A.~{Sander},
  \& H.~{Todt}, 21--26

\bibitem[{{Crowther}(2019)}]{crowther2019}
{Crowther}, P.~A. 2019, Galaxies, 7, 88

\bibitem[{{Davidson} \& {Ostriker}(1973)}]{davidson1973}
{Davidson}, K. \& {Ostriker}, J.~P. 1973, \apj, 179, 585

\bibitem[{{de Mink} \& {Belczynski}(2015)}]{demink2015}
{de Mink}, S.~E. \& {Belczynski}, K. 2015, \apj, 814, 58

\bibitem[{{de Mink} \& {Mandel}(2016)}]{selma2016}
{de Mink}, S.~E. \& {Mandel}, I. 2016, \mnras, 460, 3545

\bibitem[{{Dhawan} {et~al.}(2007){Dhawan}, {Mirabel}, {Rib{\'o}}, \&
  {Rodrigues}}]{dhawan2007}
{Dhawan}, V., {Mirabel}, I.~F., {Rib{\'o}}, M., \& {Rodrigues}, I. 2007, \apj,
  668, 430

\bibitem[{{du Buisson} {et~al.}(2020){du Buisson}, {Marchant}, {Podsiadlowski},
  {Kobayashi}, {Abdalla}, {Taylor}, {Mandel}, {de Mink}, {Moriya}, \&
  {Langer}}]{Buisson2020}
{du Buisson}, L., {Marchant}, P., {Podsiadlowski}, P., {et~al.} 2020, \mnras,
  499, 5941

\bibitem[{{Edgar}(2004)}]{Edgar2004}
{Edgar}, R. 2004, \nar, 48, 843

\bibitem[{{Ekstr{\"o}m} {et~al.}(2012){Ekstr{\"o}m}, {Georgy}, {Eggenberger},
  {Meynet}, {Mowlavi}, {Wyttenbach}, {Granada}, {Decressin}, {Hirschi},
  {Frischknecht}, {Charbonnel}, \& {Maeder}}]{ekstrom2012}
{Ekstr{\"o}m}, S., {Georgy}, C., {Eggenberger}, P., {et~al.} 2012, \aap, 537,
  A146

\bibitem[{{El Mellah}(2017)}]{elmellahthesis}
{El Mellah}, I. 2017, arXiv e-prints, arXiv:1707.09165

\bibitem[{{El Mellah} {et~al.}(2020{\natexlab{a}}){El Mellah}, {Bolte},
  {Decin}, {Homan}, \& {Keppens}}]{IEM2020}
{El Mellah}, I., {Bolte}, J., {Decin}, L., {Homan}, W., \& {Keppens}, R.
  2020{\natexlab{a}}, \aap, 637, A91

\bibitem[{{El Mellah} \& {Casse}(2017)}]{elmellah2017}
{El Mellah}, I. \& {Casse}, F. 2017, \mnras, 467, 2585

\bibitem[{{El Mellah} {et~al.}(2020{\natexlab{b}}){El Mellah}, {Grinberg},
  {Sundqvist}, {Driessen}, \& {Leutenegger}}]{ElMellah2020}
{El Mellah}, I., {Grinberg}, V., {Sundqvist}, J.~O., {Driessen}, F.~A., \&
  {Leutenegger}, M.~A. 2020{\natexlab{b}}, \aap, 643, A9

\bibitem[{{El Mellah} {et~al.}(2018){El Mellah}, {Sundqvist}, \&
  {Keppens}}]{elmellah2018}
{El Mellah}, I., {Sundqvist}, J.~O., \& {Keppens}, R. 2018, \mnras, 475, 3240

\bibitem[{{Event Horizon Telescope Collaboration} {et~al.}(2019){Event Horizon
  Telescope Collaboration}, {Akiyama}, {Alberdi}, {Alef}, {Asada}, {Azulay},
  {Baczko}, {Ball}, {Balokovi{\'c}}, {Barrett}, {Bintley}, {Blackburn},
  {Boland}, {Bouman}, {Bower}, {Bremer}, {Brinkerink}, {Brissenden}, {Britzen},
  {Broderick}, {Broguiere}, {Bronzwaer}, {Byun}, {Carlstrom}, {Chael}, {Chan},
  {Chatterjee}, {Chatterjee}, {Chen}, {Chen}, {Cho}, {Christian}, {Conway},
  {Cordes}, {Crew}, {Cui}, {Davelaar}, {De Laurentis}, {Deane}, {Dempsey},
  {Desvignes}, {Dexter}, {Doeleman}, {Eatough}, {Falcke}, {Fish}, {Fomalont},
  {Fraga-Encinas}, {Freeman}, {Friberg}, {Fromm}, {G{\'o}mez}, {Galison},
  {Gammie}, {Garc{\'\i}a}, {Gentaz}, {Georgiev}, {Goddi}, {Gold}, {Gu},
  {Gurwell}, {Hada}, {Hecht}, {Hesper}, {Ho}, {Ho}, {Honma}, {Huang}, {Huang},
  {Hughes}, {Ikeda}, {Inoue}, {Issaoun}, {James}, {Jannuzi}, {Janssen},
  {Jeter}, {Jiang}, {Johnson}, {Jorstad}, {Jung}, {Karami}, {Karuppusamy},
  {Kawashima}, {Keating}, {Kettenis}, {Kim}, {Kim}, {Kim}, {Kino}, {Koay},
  {Koch}, {Koyama}, {Kramer}, {Kramer}, {Krichbaum}, {Kuo}, {Lauer}, {Lee},
  {Li}, {Li}, {Lindqvist}, {Liu}, {Liuzzo}, {Lo}, {Lobanov}, {Loinard},
  {Lonsdale}, {Lu}, {MacDonald}, {Mao}, {Markoff}, {Marrone}, {Marscher},
  {Mart{\'\i}-Vidal}, {Matsushita}, {Matthews}, {Medeiros}, {Menten}, {Mizuno},
  {Mizuno}, {Moran}, {Moriyama}, {Moscibrodzka}, {M{\"u}ller}, {Nagai},
  {Nagar}, {Nakamura}, {Narayan}, {Narayanan}, {Natarajan}, {Neri}, {Ni},
  {Noutsos}, {Okino}, {Olivares}, {Ortiz-Le{\'o}n}, {Oyama}, {{\"O}zel},
  {Palumbo}, {Patel}, {Pen}, {Pesce}, {Pi{\'e}tu}, {Plambeck}, {PopStefanija},
  {Porth}, {Prather}, {Preciado-L{\'o}pez}, {Psaltis}, {Pu}, {Ramakrishnan},
  {Rao}, {Rawlings}, {Raymond}, {Rezzolla}, {Ripperda}, {Roelofs}, {Rogers},
  {Ros}, {Rose}, {Roshanineshat}, {Rottmann}, {Roy}, {Ruszczyk}, {Ryan},
  {Rygl}, {S{\'a}nchez}, {S{\'a}nchez-Arguelles}, {Sasada}, {Savolainen},
  {Schloerb}, {Schuster}, {Shao}, {Shen}, {Small}, {Sohn}, {SooHoo}, {Tazaki},
  {Tiede}, {Tilanus}, {Titus}, {Toma}, {Torne}, {Trent}, {Trippe}, {Tsuda},
  {van Bemmel}, {van Langevelde}, {van Rossum}, {Wagner}, {Wardle},
  {Weintroub}, {Wex}, {Wharton}, {Wielgus}, {Wong}, {Wu}, {Young}, {Young},
  {Younsi}, {Yuan}, {Yuan}, {Zensus}, {Zhao}, {Zhao}, {Zhu}, {Algaba},
  {Allardi}, {Amestica}, {Anczarski}, {Bach}, {Baganoff}, {Beaudoin}, {Benson},
  {Berthold}, {Blanchard}, {Blundell}, {Bustamente}, {Cappallo},
  {Castillo-Dom{\'\i}nguez}, {Chang}, {Chang}, {Chang}, {Chen}, {Chilson},
  {Chuter}, {C{\'o}rdova Rosado}, {Coulson}, {Crawford}, {Crowley}, {David},
  {Derome}, {Dexter}, {Dornbusch}, {Dudevoir}, {Dzib}, {Eckart}, {Eckert},
  {Erickson}, {Everett}, {Faber}, {Farah}, {Fath}, {Folkers}, {Forbes},
  {Freund}, {G{\'o}mez-Ruiz}, {Gale}, {Gao}, {Geertsema}, {Graham}, {Greer},
  {Grosslein}, {Gueth}, {Haggard}, {Halverson}, {Han}, {Han}, {Hao},
  {Hasegawa}, {Henning}, {Hern{\'a}ndez-G{\'o}mez}, {Herrero-Illana},
  {Heyminck}, {Hirota}, {Hoge}, {Huang}, {Impellizzeri}, {Jiang}, {Kamble},
  {Keisler}, {Kimura}, {Kono}, {Kubo}, {Kuroda}, {Lacasse}, {Laing}, {Leitch},
  {Li}, {Lin}, {Liu}, {Liu}, {Lu}, {Marson}, {Martin-Cocher}, {Massingill},
  {Matulonis}, {McColl}, {McWhirter}, {Messias}, {Meyer-Zhao}, {Michalik},
  {Monta{\~n}a}, {Montgomerie}, {Mora-Klein}, {Muders}, {Nadolski}, {Navarro},
  {Neilsen}, {Nguyen}, {Nishioka}, {Norton}, {Nowak}, {Nystrom}, {Ogawa},
  {Oshiro}, {Oyama}, {Parsons}, {Paine}, {Pe{\~n}alver}, {Phillips}, {Poirier},
  {Pradel}, {Primiani}, {Raffin}, {Rahlin}, {Reiland}, {Risacher}, {Ruiz},
  {S{\'a}ez-Mada{\'\i}n}, {Sassella}, {Schellart}, {Shaw}, {Silva}, {Shiokawa},
  {Smith}, {Snow}, {Souccar}, {Sousa}, {Sridharan}, {Srinivasan}, {Stahm},
  {Stark}, {Story}, {Timmer}, {Vertatschitsch}, {Walther}, {Wei}, {Whitehorn},
  {Whitney}, {Woody}, {Wouterloot}, {Wright}, {Yamaguchi}, {Yu}, {Zeballos},
  {Zhang}, \& {Ziurys}}]{eht2019}
{Event Horizon Telescope Collaboration}, {Akiyama}, K., {Alberdi}, A., {et~al.}
  2019, \apjl, 875, L1

\bibitem[{{Farr} {et~al.}(2011){Farr}, {Sravan}, {Cantrell}, {Kreidberg},
  {Bailyn}, {Mandel}, \& {Kalogera}}]{farr2011}
{Farr}, W.~M., {Sravan}, N., {Cantrell}, A., {et~al.} 2011, \apj, 741, 103

\bibitem[{{Feldmeier}(1995)}]{Feldmeier1995}
{Feldmeier}, A. 1995, \aap, 299, 523

\bibitem[{{Fragos} {et~al.}(2009){Fragos}, {Willems}, {Kalogera}, {Ivanova},
  {Rockefeller}, {Fryer}, \& {Young}}]{fragos2009}
{Fragos}, T., {Willems}, B., {Kalogera}, V., {et~al.} 2009, \apj, 697, 1057

\bibitem[{{Frank} {et~al.}(2002){Frank}, {King}, \& {Raine}}]{FKR2002}
{Frank}, J., {King}, A., \& {Raine}, D.~J. 2002, {Accretion Power in
  Astrophysics: Third Edition}

\bibitem[{{Fujita} {et~al.}(1998){Fujita}, {Inoue}, {Nakamura}, {Manmoto}, \&
  {Nakamura}}]{fujita1998}
{Fujita}, Y., {Inoue}, S., {Nakamura}, T., {Manmoto}, T., \& {Nakamura}, K.~E.
  1998, \apjl, 495, L85

\bibitem[{{F{\"u}rst} {et~al.}(2016){F{\"u}rst}, {Walton}, {Harrison}, {Stern},
  {Barret}, {Brightman}, {Fabian}, {Grefenstette}, {Madsen}, {Middleton},
  {Miller}, {Pottschmidt}, {Ptak}, {Rana}, \& {Webb}}]{Fuerst2016}
{F{\"u}rst}, F., {Walton}, D.~J., {Harrison}, F.~A., {et~al.} 2016, \apjl, 831,
  L14

\bibitem[{{Gou} {et~al.}(2011){Gou}, {McClintock}, {Reid}, {Orosz}, {Steiner},
  {Narayan}, {Xiang}, {Remillard}, {Arnaud}, \& {Davis}}]{gou2011}
{Gou}, L., {McClintock}, J.~E., {Reid}, M.~J., {et~al.} 2011, \apj, 742, 85

\bibitem[{{Grassitelli} {et~al.}(2015){Grassitelli}, {Fossati},
  {Sim{\'o}n-Di{\'a}z}, {Langer}, {Castro}, \& {Sanyal}}]{Grassitelli2015}
{Grassitelli}, L., {Fossati}, L., {Sim{\'o}n-Di{\'a}z}, S., {et~al.} 2015,
  \apjl, 808, L31

\bibitem[{{Grinberg} {et~al.}(2017){Grinberg}, {Hell}, {El Mellah}, {Neilsen},
  {Sander}, {Leutenegger}, {F{\"u}rst}, {Huenemoerder}, {Kretschmar},
  {K{\"u}hnel}, {Mart{\'\i}nez-N{\'u}{\~n}ez}, {Niu}, {Pottschmidt}, {Schulz},
  {Wilms}, \& {Nowak}}]{Grinberg2017}
{Grinberg}, V., {Hell}, N., {El Mellah}, I., {et~al.} 2017, \aap, 608, A143

\bibitem[{{Groenewegen} \& {Lamers}(1989)}]{groenewegen1989}
{Groenewegen}, M.~A.~T. \& {Lamers}, H.~J.~G.~L.~M. 1989, \aaps, 79, 359

\bibitem[{{Hirsch} {et~al.}(2019){Hirsch}, {Hell}, {Grinberg}, {Ballhausen},
  {Nowak}, {Pottschmidt}, {Schulz}, {Dauser}, {Hanke}, {Kallman}, {Brown}, \&
  {Wilms}}]{Hirsch19}
{Hirsch}, M., {Hell}, N., {Grinberg}, V., {et~al.} 2019, \aap, 626, A64

\bibitem[{{Hobbs} {et~al.}(2005){Hobbs}, {Lorimer}, {Lyne}, \&
  {Kramer}}]{hobbs2005}
{Hobbs}, G., {Lorimer}, D.~R., {Lyne}, A.~G., \& {Kramer}, M. 2005, \mnras,
  360, 974

\bibitem[{{Hudec} {et~al.}(2007){Hudec}, {Pina}, {{\v{S}}imon},
  {{\v{S}}v{\'e}da}, {Inneman}, {Semencov{\'a}}, \&
  {Skulinov{\'a}}}]{Hudec2007}
{Hudec}, R., {Pina}, L., {{\v{S}}imon}, V., {et~al.} 2007, Nuclear Physics B
  Proceedings Supplements, 166, 229

\bibitem[{{Iben} \& {Tutukov}(1996)}]{iben1996}
{Iben}, Icko, J. \& {Tutukov}, A.~V. 1996, \apjs, 105, 145

\bibitem[{{Illarionov} \& {Sunyaev}(1975)}]{Illarionov1975}
{Illarionov}, A.~F. \& {Sunyaev}, R.~A. 1975, \aap, 39, 185

\bibitem[{{In't Zand} {et~al.}(1994){In't Zand}, {Priedhorsky}, {Moss},
  {Fenimore}, {Black}, {Kelley}, {Stilwell}, {Birsa}, {Borozdin}, \&
  {Arefiev}}]{zand1994}
{In't Zand}, J.~J., {Priedhorsky}, W.~C., {Moss}, C.~E., {et~al.} 1994, in
  Society of Photo-Optical Instrumentation Engineers (SPIE) Conference Series,
  Vol. 2279, Advances in Multilayer and Grazing Incidence X-Ray/EUV/FUV Optics,
  ed. R.~B. {Hoover} \& A.~B. {Walker}, 458--468

\bibitem[{{Israel} {et~al.}(2017){Israel}, {Belfiore}, {Stella}, {Esposito},
  {Casella}, {De Luca}, {Marelli}, {Papitto}, {Perri}, {Puccetti}, {Castillo},
  {Salvetti}, {Tiengo}, {Zampieri}, {D'Agostino}, {Greiner}, {Haberl},
  {Novara}, {Salvaterra}, {Turolla}, {Watson}, {Wilms}, \&
  {Wolter}}]{Israel2017}
{Israel}, G.~L., {Belfiore}, A., {Stella}, L., {et~al.} 2017, Science, 355, 817

\bibitem[{{King}(2008)}]{king2008}
{King}, A.~R. 2008, \mnras, 385, L113

\bibitem[{{Kremer} {et~al.}(2019){Kremer}, {Lu}, {Rodriguez}, {Lachat}, \&
  {Rasio}}]{kremer2019}
{Kremer}, K., {Lu}, W., {Rodriguez}, C.~L., {Lachat}, M., \& {Rasio}, F.~A.
  2019, \apj, 881, 75

\bibitem[{{Kretschmar} {et~al.}(2021){Kretschmar}, {El Mellah},
  {Mart{\'\i}nez-N{\'u}{\~n}ez}, {F{\"u}rst}, {Grinberg}, {Sander}, {van den
  Eijnden}, {Degenaar}, {Ma{\'\i}z-Apell{\'a}niz}, {Jim{\'e}nez Esteban},
  {Ramos-Lerate}, \& {Utrilla}}]{Kretschmar2021}
{Kretschmar}, P., {El Mellah}, I., {Mart{\'\i}nez-N{\'u}{\~n}ez}, S., {et~al.}
  2021, arXiv e-prints, arXiv:2104.13148

\bibitem[{{Kruckow} {et~al.}(2018){Kruckow}, {Tauris}, {Langer}, {Kramer}, \&
  {Izzard}}]{kruckow2018}
{Kruckow}, M.~U., {Tauris}, T.~M., {Langer}, N., {Kramer}, M., \& {Izzard},
  R.~G. 2018, \mnras, 481, 1908

\bibitem[{{Lamers} {et~al.}(1995){Lamers}, {Snow}, \& {Lindholm}}]{lamers1995}
{Lamers}, H. J.~G.~L.~M., {Snow}, T.~P., \& {Lindholm}, D.~M. 1995, \apj, 455,
  269

\bibitem[{{Langer}(1992)}]{langer1992}
{Langer}, N. 1992, \aap, 265, L17

\bibitem[{{Langer}(2012)}]{langer2012}
{Langer}, N. 2012, \araa, 50, 107

\bibitem[{{Langer} {et~al.}(2020){Langer}, {Sch{\"u}rmann}, {Stoll},
  {Marchant}, {Lennon}, {Mahy}, {de Mink}, {Quast}, {Riedel}, {Sana},
  {Schneider}, {Schootemeijer}, {Wang}, {Almeida}, {Bestenlehner},
  {Bodensteiner}, {Castro}, {Clark}, {Crowther}, {Dufton}, {Evans}, {Fossati},
  {Gr{\"a}fener}, {Grassitelli}, {Grin}, {Hastings}, {Herrero}, {de Koter},
  {Menon}, {Patrick}, {Puls}, {Renzo}, {Sander}, {Schneider}, {Sen}, {Shenar},
  {Sim{\'o}n-D{\'\i}as}, {Tauris}, {Tramper}, {Vink}, \& {Xu}}]{langer2020}
{Langer}, N., {Sch{\"u}rmann}, C., {Stoll}, K., {et~al.} 2020, \aap, 638, A39

\bibitem[{{Laplace} {et~al.}(2020){Laplace}, {G{\"o}tberg}, {de Mink},
  {Justham}, \& {Farmer}}]{laplace2020}
{Laplace}, E., {G{\"o}tberg}, Y., {de Mink}, S.~E., {Justham}, S., \& {Farmer},
  R. 2020, \aap, 637, A6

\bibitem[{{Laplace} {et~al.}(2021){Laplace}, {Justham}, {Renzo}, {G{\"o}tberg},
  {Farmer}, {Vartanyan}, \& {de Mink}}]{laplace2021}
{Laplace}, E., {Justham}, S., {Renzo}, M., {et~al.} 2021, arXiv e-prints,
  arXiv:2102.05036

\bibitem[{{Liao} {et~al.}(2020){Liao}, {Liu}, {Zheng}, \& {Gou}}]{Liao2020}
{Liao}, Z., {Liu}, J., {Zheng}, X., \& {Gou}, L. 2020, \mnras, 492, 5922

\bibitem[{{Livio} {et~al.}(1986){Livio}, {Soker}, {de Kool}, \&
  {Savonije}}]{livio1986}
{Livio}, M., {Soker}, N., {de Kool}, M., \& {Savonije}, G.~J. 1986, \mnras,
  222, 235

\bibitem[{{Mandel} \& {M{\"u}ller}(2020)}]{mandel2020a}
{Mandel}, I. \& {M{\"u}ller}, B. 2020, \mnras, 499, 3214

\bibitem[{{Mandel} {et~al.}(2021){Mandel}, {M{\"u}ller}, {Riley}, {de Mink},
  {Vigna-G{\'o}mez}, \& {Chattopadhyay}}]{mandel2020b}
{Mandel}, I., {M{\"u}ller}, B., {Riley}, J., {et~al.} 2021, \mnras, 500, 1380

\bibitem[{{Manousakis} {et~al.}(2012){Manousakis}, {Walter}, \&
  {Blondin}}]{Manousakis2012}
{Manousakis}, A., {Walter}, R., \& {Blondin}, J.~M. 2012, \aap, 547, A20

\bibitem[{{Marchant} {et~al.}(2016){Marchant}, {Langer}, {Podsiadlowski},
  {Tauris}, \& {Moriya}}]{Marchant2016}
{Marchant}, P., {Langer}, N., {Podsiadlowski}, P., {Tauris}, T.~M., \&
  {Moriya}, T.~J. 2016, \aap, 588, A50

\bibitem[{{Mashian} \& {Loeb}(2017)}]{Mashian2017}
{Mashian}, N. \& {Loeb}, A. 2017, \mnras, 470, 2611

\bibitem[{{Masuda} \& {Hotokezaka}(2019)}]{masuda2019}
{Masuda}, K. \& {Hotokezaka}, K. 2019, \apj, 883, 169

\bibitem[{{Mennekens} \& {Vanbeveren}(2014)}]{mennekens2014}
{Mennekens}, N. \& {Vanbeveren}, D. 2014, \aap, 564, A134

\bibitem[{{Merloni} {et~al.}(2012){Merloni}, {Predehl}, {Becker},
  {B{\"o}hringer}, {Boller}, {Brunner}, {Brusa}, {Dennerl}, {Freyberg},
  {Friedrich}, {Georgakakis}, {Haberl}, {Hasinger}, {Meidinger}, {Mohr},
  {Nandra}, {Rau}, {Reiprich}, {Robrade}, {Salvato}, {Santangelo}, {Sasaki},
  {Schwope}, {Wilms}, \& {German eROSITA Consortium}}]{Merloni2012}
{Merloni}, A., {Predehl}, P., {Becker}, W., {et~al.} 2012, arXiv e-prints,
  arXiv:1209.3114

\bibitem[{{Miller-Jones} {et~al.}(2021){Miller-Jones}, {Bahramian}, {Orosz},
  {Mandel}, {Gou}, {Maccarone}, {Neijssel}, {Zhao}, {Zi{\'o}{\l}kowski},
  {Reid}, {Uttley}, {Zheng}, {Byun}, {Dodson}, {Grinberg}, {Jung}, {Kim},
  {Marcote}, {Markoff}, {Rioja}, {Rushton}, {Russell}, {Sivakoff}, {Tetarenko},
  {Tudose}, \& {Wilms}}]{miller-jones2021}
{Miller-Jones}, J. C.~A., {Bahramian}, A., {Orosz}, J.~A., {et~al.} 2021,
  Science, 371, 1046

\bibitem[{{Minniti} {et~al.}(2015){Minniti}, {Contreras Ramos},
  {Alonso-Garc{\'\i}a}, {Anguita}, {Catelan}, {Gran}, {Motta}, {Muro}, {Rojas},
  \& {Saito}}]{minniti2015}
{Minniti}, D., {Contreras Ramos}, R., {Alonso-Garc{\'\i}a}, J., {et~al.} 2015,
  \apjl, 810, L20

\bibitem[{{Mirabel} \& {Rodrigues}(2003)}]{mirabel2003}
{Mirabel}, I.~F. \& {Rodrigues}, I. 2003, Science, 300, 1119

\bibitem[{{Narayan} \& {McClintock}(2008)}]{Narayan2008}
{Narayan}, R. \& {McClintock}, J.~E. 2008, \nar, 51, 733

\bibitem[{{Narayan} \& {Yi}(1994)}]{Narayan1994}
{Narayan}, R. \& {Yi}, I. 1994, \apjl, 428, L13

\bibitem[{{Narayan} \& {Yi}(1995{\natexlab{a}})}]{Narayan1995a}
{Narayan}, R. \& {Yi}, I. 1995{\natexlab{a}}, \apj, 444, 231

\bibitem[{{Narayan} \& {Yi}(1995{\natexlab{b}})}]{Narayan1995b}
{Narayan}, R. \& {Yi}, I. 1995{\natexlab{b}}, \apj, 452, 710

\bibitem[{{Neijssel} {et~al.}(2021){Neijssel}, {Vinciguerra},
  {Vigna-G{\'o}mez}, {Hirai}, {Miller-Jones}, {Bahramian}, {Maccarone}, \&
  {Mandel}}]{neijssel2021}
{Neijssel}, C.~J., {Vinciguerra}, S., {Vigna-G{\'o}mez}, A., {et~al.} 2021,
  \apj, 908, 118

\bibitem[{{Novikov} \& {Thorne}(1973)}]{Novikov1973}
{Novikov}, I.~D. \& {Thorne}, K.~S. 1973, in Black Holes (Les Astres Occlus),
  343--450

\bibitem[{{O'Connor} \& {Ott}(2011)}]{oconner2011}
{O'Connor}, E. \& {Ott}, C.~D. 2011, \apj, 730, 70

\bibitem[{{Orosz} {et~al.}(2011){Orosz}, {McClintock}, {Aufdenberg},
  {Remillard}, {Reid}, {Narayan}, \& {Gou}}]{orosz2011}
{Orosz}, J.~A., {McClintock}, J.~E., {Aufdenberg}, J.~P., {et~al.} 2011, \apj,
  742, 84

\bibitem[{{O'Shaughnessy} {et~al.}(2008){O'Shaughnessy}, {Kim}, {Kalogera}, \&
  {Belczynski}}]{shaughnessy2008}
{O'Shaughnessy}, R., {Kim}, C., {Kalogera}, V., \& {Belczynski}, K. 2008, \apj,
  672, 479

\bibitem[{{Owocki} {et~al.}(1988){Owocki}, {Castor}, \& {Rybicki}}]{Owocki1988}
{Owocki}, S.~P., {Castor}, J.~I., \& {Rybicki}, G.~B. 1988, \apj, 335, 914

\bibitem[{{Owocki} \& {Rybicki}(1984)}]{Owocki1984}
{Owocki}, S.~P. \& {Rybicki}, G.~B. 1984, \apj, 284, 337

\bibitem[{{{\"O}zel} {et~al.}(2010){{\"O}zel}, {Psaltis}, {Narayan}, \&
  {McClintock}}]{ozel2010}
{{\"O}zel}, F., {Psaltis}, D., {Narayan}, R., \& {McClintock}, J.~E. 2010,
  \apj, 725, 1918

\bibitem[{{Packet}(1981)}]{packet1981}
{Packet}, W. 1981, \aap, 102, 17

\bibitem[{{Perets} {et~al.}(2016){Perets}, {Li}, {Lombardi}, \&
  {Milcarek}}]{perets2016}
{Perets}, H.~B., {Li}, Z., {Lombardi}, James~C., J., \& {Milcarek}, Stephen~R.,
  J. 2016, \apj, 823, 113

\bibitem[{{Petrovic} {et~al.}(2005{\natexlab{a}}){Petrovic}, {Langer}, \& {van
  der Hucht}}]{Petrovic2005a}
{Petrovic}, J., {Langer}, N., \& {van der Hucht}, K.~A. 2005{\natexlab{a}},
  \aap, 435, 1013

\bibitem[{{Petrovic} {et~al.}(2005{\natexlab{b}}){Petrovic}, {Langer}, {Yoon},
  \& {Heger}}]{Petrovic2005b}
{Petrovic}, J., {Langer}, N., {Yoon}, S.~C., \& {Heger}, A. 2005{\natexlab{b}},
  \aap, 435, 247

\bibitem[{{Priedhorsky} {et~al.}(1996){Priedhorsky}, {Peele}, \&
  {Nugent}}]{Priedhorsky1996}
{Priedhorsky}, W.~C., {Peele}, A.~G., \& {Nugent}, K.~A. 1996, \mnras, 279, 733

\bibitem[{{Puls} {et~al.}(1996){Puls}, {Kudritzki}, {Herrero}, {Pauldrach},
  {Haser}, {Lennon}, {Gabler}, {Voels}, {Vilchez}, {Wachter}, \&
  {Feldmeier}}]{puls1996}
{Puls}, J., {Kudritzki}, R.~P., {Herrero}, A., {et~al.} 1996, \aap, 305, 171

\bibitem[{{Qin} {et~al.}(2018){Qin}, {Fragos}, {Meynet}, {Andrews},
  {S{\o}rensen}, \& {Song}}]{qin2018}
{Qin}, Y., {Fragos}, T., {Meynet}, G., {et~al.} 2018, \aap, 616, A28

\bibitem[{Qin {et~al.}(2019)Qin, Marchant, Fragos, Meynet, \&
  Kalogera}]{qin2019}
Qin, Y., Marchant, P., Fragos, T., Meynet, G., \& Kalogera, V. 2019, The
  Astrophysical Journal, 870, L18

\bibitem[{{Quast} {et~al.}(2019){Quast}, {Langer}, \& {Tauris}}]{quast2019}
{Quast}, M., {Langer}, N., \& {Tauris}, T.~M. 2019, \aap, 628, A19

\bibitem[{{Reig}(2011)}]{Reig2011}
{Reig}, P. 2011, \apss, 332, 1

\bibitem[{{Repetto} {et~al.}(2012){Repetto}, {Davies}, \&
  {Sigurdsson}}]{repetto2012}
{Repetto}, S., {Davies}, M.~B., \& {Sigurdsson}, S. 2012, \mnras, 425, 2799

\bibitem[{{Repetto} {et~al.}(2017){Repetto}, {Igoshev}, \&
  {Nelemans}}]{repetto2017}
{Repetto}, S., {Igoshev}, A.~P., \& {Nelemans}, G. 2017, \mnras, 467, 298

\bibitem[{{Rosslowe} \& {Crowther}(2015)}]{rosslowe2015}
{Rosslowe}, C.~K. \& {Crowther}, P.~A. 2015, \mnras, 447, 2322

\bibitem[{{Ruffert}(1999)}]{ruffert1999}
{Ruffert}, M. 1999, \aap, 346, 861

\bibitem[{{Scarcella} {et~al.}(2020){Scarcella}, {Gaggero}, {Connors},
  {Manshanden}, {Ricotti}, \& {Bertone}}]{Scarcella20}
{Scarcella}, F., {Gaggero}, D., {Connors}, R., {et~al.} 2020, arXiv e-prints,
  arXiv:2012.10421

\bibitem[{{Schweickhardt} {et~al.}(1999){Schweickhardt}, {Schmutz}, {Stahl},
  {Szeifert}, \& {Wolf}}]{Schweickhardt1999}
{Schweickhardt}, J., {Schmutz}, W., {Stahl}, O., {Szeifert}, T., \& {Wolf}, B.
  1999, \aap, 347, 127

\bibitem[{{Shakura} \& {Sunyaev}(1973)}]{Shakura1973}
{Shakura}, N.~I. \& {Sunyaev}, R.~A. 1973, \aap, 500, 33

\bibitem[{Shao \& Li(2020)}]{shao2020}
Shao, Y. \& Li, X.-D. 2020, The Astrophysical Journal, 898, 143

\bibitem[{{Shapiro} \& {Lightman}(1976)}]{shapiro1976}
{Shapiro}, S.~L. \& {Lightman}, A.~P. 1976, \apj, 204, 555

\bibitem[{{Shapiro} \& {Teukolsky}(1983)}]{shapiro1983}
{Shapiro}, S.~L. \& {Teukolsky}, S.~A. 1983, {Black holes, white dwarfs, and
  neutron stars : the physics of compact objects}

\bibitem[{{Shara} {et~al.}(2020){Shara}, {Crawford}, {Vanbeveren}, {Moffat},
  {Zurek}, \& {Crause}}]{shara2020}
{Shara}, M.~M., {Crawford}, S.~M., {Vanbeveren}, D., {et~al.} 2020, \mnras,
  492, 4430

\bibitem[{{Soker} {et~al.}(1986){Soker}, {Livio}, {de Kool}, \&
  {Savonije}}]{soker1986}
{Soker}, N., {Livio}, M., {de Kool}, M., \& {Savonije}, G.~J. 1986, \mnras,
  221, 445

\bibitem[{{Stevenson} {et~al.}(2015){Stevenson}, {Ohme}, \&
  {Fairhurst}}]{stevenson2015}
{Stevenson}, S., {Ohme}, F., \& {Fairhurst}, S. 2015, \apj, 810, 58

\bibitem[{{Sukhbold} {et~al.}(2018){Sukhbold}, {Woosley}, \&
  {Heger}}]{sukhbold2018}
{Sukhbold}, T., {Woosley}, S.~E., \& {Heger}, A. 2018, \apj, 860, 93

\bibitem[{{Sundqvist} {et~al.}(2018){Sundqvist}, {Owocki}, \&
  {Puls}}]{Sundqvist2018}
{Sundqvist}, J.~O., {Owocki}, S.~P., \& {Puls}, J. 2018, \aap, 611, A17

\bibitem[{{Tsuna} {et~al.}(2018){Tsuna}, {Kawanaka}, \& {Totani}}]{tsuna2018}
{Tsuna}, D., {Kawanaka}, N., \& {Totani}, T. 2018, \mnras, 477, 791

\bibitem[{{Ugliano} {et~al.}(2012){Ugliano}, {Janka}, {Marek}, \&
  {Arcones}}]{ugliano2012}
{Ugliano}, M., {Janka}, H.-T., {Marek}, A., \& {Arcones}, A. 2012, \apj, 757,
  69

\bibitem[{{van den Heuvel} {et~al.}(2017){van den Heuvel}, {Portegies Zwart},
  \& {de Mink}}]{Heuvel2017}
{van den Heuvel}, E.~P.~J., {Portegies Zwart}, S.~F., \& {de Mink}, S.~E. 2017,
  \mnras, 471, 4256

\bibitem[{{van der Hucht}(2001)}]{hucht2001}
{van der Hucht}, K.~A. 2001, VizieR Online Data Catalog, III/215

\bibitem[{{van der Hucht}(2006)}]{hucht2006}
{van der Hucht}, K.~A. 2006, \aap, 458, 453

\bibitem[{{Vanbeveren} {et~al.}(1998{\natexlab{a}}){Vanbeveren}, {De Donder},
  {Van Bever}, {Van Rensbergen}, \& {De Loore}}]{vanbeveren1998b}
{Vanbeveren}, D., {De Donder}, E., {Van Bever}, J., {Van Rensbergen}, W., \&
  {De Loore}, C. 1998{\natexlab{a}}, \na, 3, 443

\bibitem[{{Vanbeveren} {et~al.}(1998{\natexlab{b}}){Vanbeveren}, {De Loore}, \&
  {Van Rensbergen}}]{vanbeveren1998a}
{Vanbeveren}, D., {De Loore}, C., \& {Van Rensbergen}, W. 1998{\natexlab{b}},
  \aapr, 9, 63

\bibitem[{{Vanbeveren} {et~al.}(2018){Vanbeveren}, {Mennekens}, {Shara}, \&
  {Moffat}}]{vanbeveren2018}
{Vanbeveren}, D., {Mennekens}, N., {Shara}, M.~M., \& {Moffat}, A.~F.~J. 2018,
  \aap, 615, A65

\bibitem[{{Vanbeveren} {et~al.}(2020){Vanbeveren}, {Mennekens}, {van den
  Heuvel}, \& {Van Bever}}]{vanbeveren2020}
{Vanbeveren}, D., {Mennekens}, N., {van den Heuvel}, E.~P.~J., \& {Van Bever},
  J. 2020, \aap, 636, A99

\bibitem[{{Vilhu} {et~al.}(2021){Vilhu}, {Kallman}, {Koljonen}, \&
  {Hannikainen}}]{Vilhu2021}
{Vilhu}, O., {Kallman}, T.~R., {Koljonen}, K.~I., \& {Hannikainen}, D.~C. 2021,
  arXiv e-prints, arXiv:2104.02305

\bibitem[{{Vink} {et~al.}(2001){Vink}, {de Koter}, \& {Lamers}}]{vink2001}
{Vink}, J.~S., {de Koter}, A., \& {Lamers}, H.~J.~G.~L.~M. 2001, \aap, 369, 574

\bibitem[{{Vink} \& {Sander}(2021)}]{vink2021}
{Vink}, J.~S. \& {Sander}, A. A.~C. 2021, \mnras, 504, 2051

\bibitem[{{Waters} {et~al.}(1988){Waters}, {van den Heuvel}, {Taylor},
  {Habets}, \& {Persi}}]{waters1988}
{Waters}, L.~B.~F.~M., {van den Heuvel}, E.~P.~J., {Taylor}, A.~R., {Habets},
  G.~M.~H.~J., \& {Persi}, P. 1988, \aap, 198, 200

\bibitem[{{Wellstein} \& {Langer}(1999)}]{Wellstein1999}
{Wellstein}, S. \& {Langer}, N. 1999, \aap, 350, 148

\bibitem[{{Wiktorowicz} {et~al.}(2020){Wiktorowicz}, {Lu}, {Wyrzykowski},
  {Zhang}, {Liu}, {Justham}, \& {Belczynski}}]{wiktorowicz2020}
{Wiktorowicz}, G., {Lu}, Y., {Wyrzykowski}, {\L}., {et~al.} 2020, \apj, 905,
  134

\bibitem[{{Willems} {et~al.}(2005){Willems}, {Henninger}, {Levin}, {Ivanova},
  {Kalogera}, {McGhee}, {Timmes}, \& {Fryer}}]{willems2005}
{Willems}, B., {Henninger}, M., {Levin}, T., {et~al.} 2005, \apj, 625, 324

\bibitem[{{Wong} {et~al.}(2012){Wong}, {Valsecchi}, {Fragos}, \&
  {Kalogera}}]{wong2012}
{Wong}, T.-W., {Valsecchi}, F., {Fragos}, T., \& {Kalogera}, V. 2012, \apj,
  747, 111

\bibitem[{{Wood} {et~al.}(1984){Wood}, {Meekins}, {Yentis}, {Smathers},
  {McNutt}, {Bleach}, {Byram}, {Chupp}, {Friedman}, \& {Meidav}}]{wood1984}
{Wood}, K.~S., {Meekins}, J.~F., {Yentis}, D.~J., {et~al.} 1984, \apjs, 56, 507

\bibitem[{{Woosley}(2019)}]{woosley2019}
{Woosley}, S.~E. 2019, \apj, 878, 49

\bibitem[{{Woosley} {et~al.}(2020){Woosley}, {Sukhbold}, \&
  {Janka}}]{woosley2020}
{Woosley}, S.~E., {Sukhbold}, T., \& {Janka}, H.~T. 2020, \apj, 896, 56

\bibitem[{{Wyrzykowski} \& {Mandel}(2020)}]{wyrzykowski2020}
{Wyrzykowski}, {\L}. \& {Mandel}, I. 2020, \aap, 636, A20

\bibitem[{{Yalinewich} {et~al.}(2018){Yalinewich}, {Beniamini}, {Hotokezaka},
  \& {Zhu}}]{Yalinewich2018}
{Yalinewich}, A., {Beniamini}, P., {Hotokezaka}, K., \& {Zhu}, W. 2018, \mnras,
  481, 930

\bibitem[{{Yamaguchi} {et~al.}(2018){Yamaguchi}, {Kawanaka}, {Bulik}, \&
  {Piran}}]{Yamaguchi2018}
{Yamaguchi}, M.~S., {Kawanaka}, N., {Bulik}, T., \& {Piran}, T. 2018, \apj,
  861, 21

\bibitem[{{Yuan} \& {Narayan}(2014)}]{Yuan2014}
{Yuan}, F. \& {Narayan}, R. 2014, \araa, 52, 529

\bibitem[{{Zhao} {et~al.}(2021){Zhao}, {Gou}, {Dong}, {Zheng}, {Steiner},
  {Miller-Jones}, {Bahramian}, {Orosz}, \& {Feng}}]{Zhao2021}
{Zhao}, X., {Gou}, L., {Dong}, Y., {et~al.} 2021, \apj, 908, 117

\bibitem[{{Zucker} {et~al.}(2007){Zucker}, {Mazeh}, \&
  {Alexander}}]{zucker2007}
{Zucker}, S., {Mazeh}, T., \& {Alexander}, T. 2007, \apj, 670, 1326

\end{thebibliography}

\begin{appendix}

\section{Mass accretion rate}

\begin{figure}
    \centering
    \includegraphics[width=\linewidth]{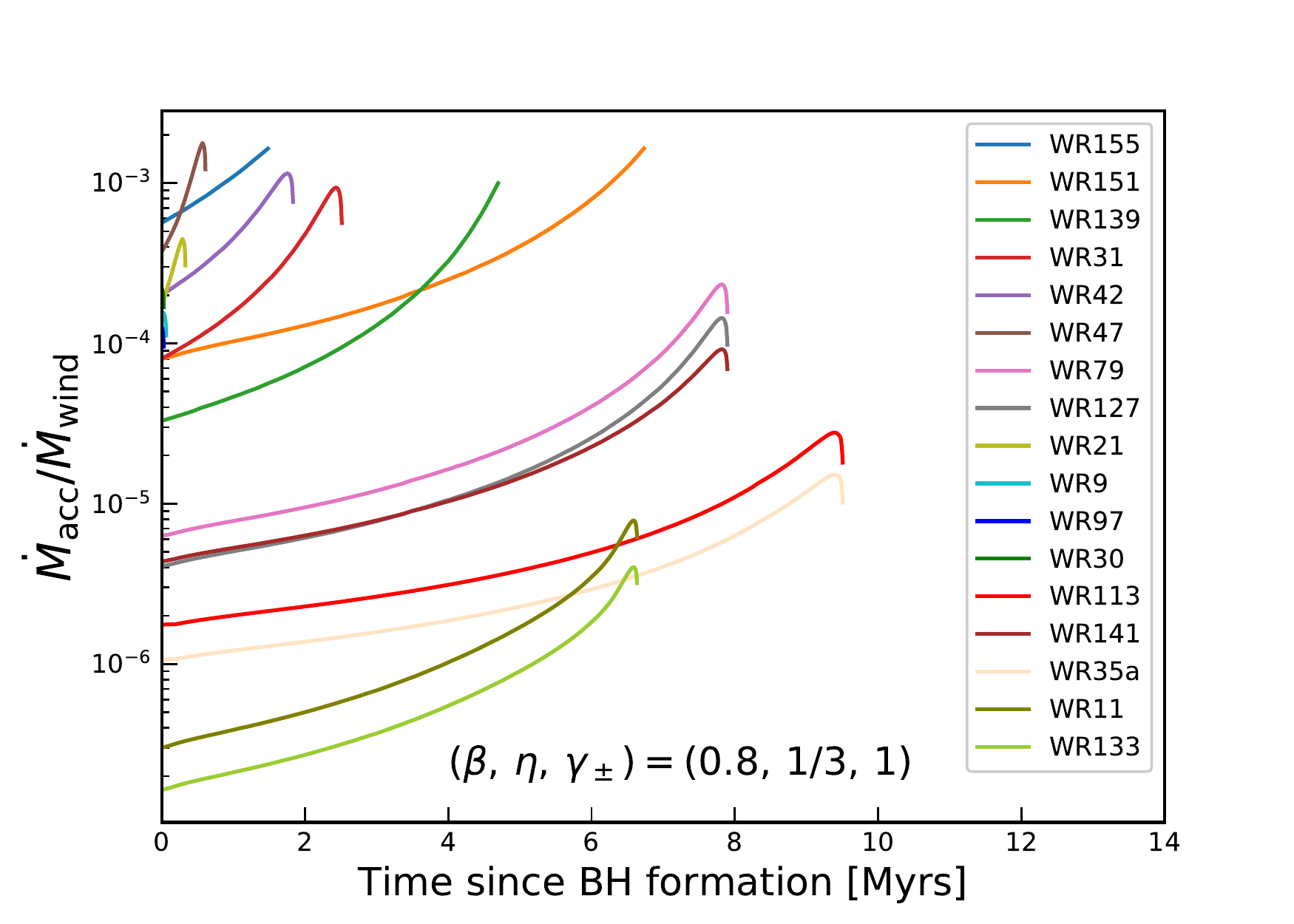}
    \includegraphics[width=\linewidth]{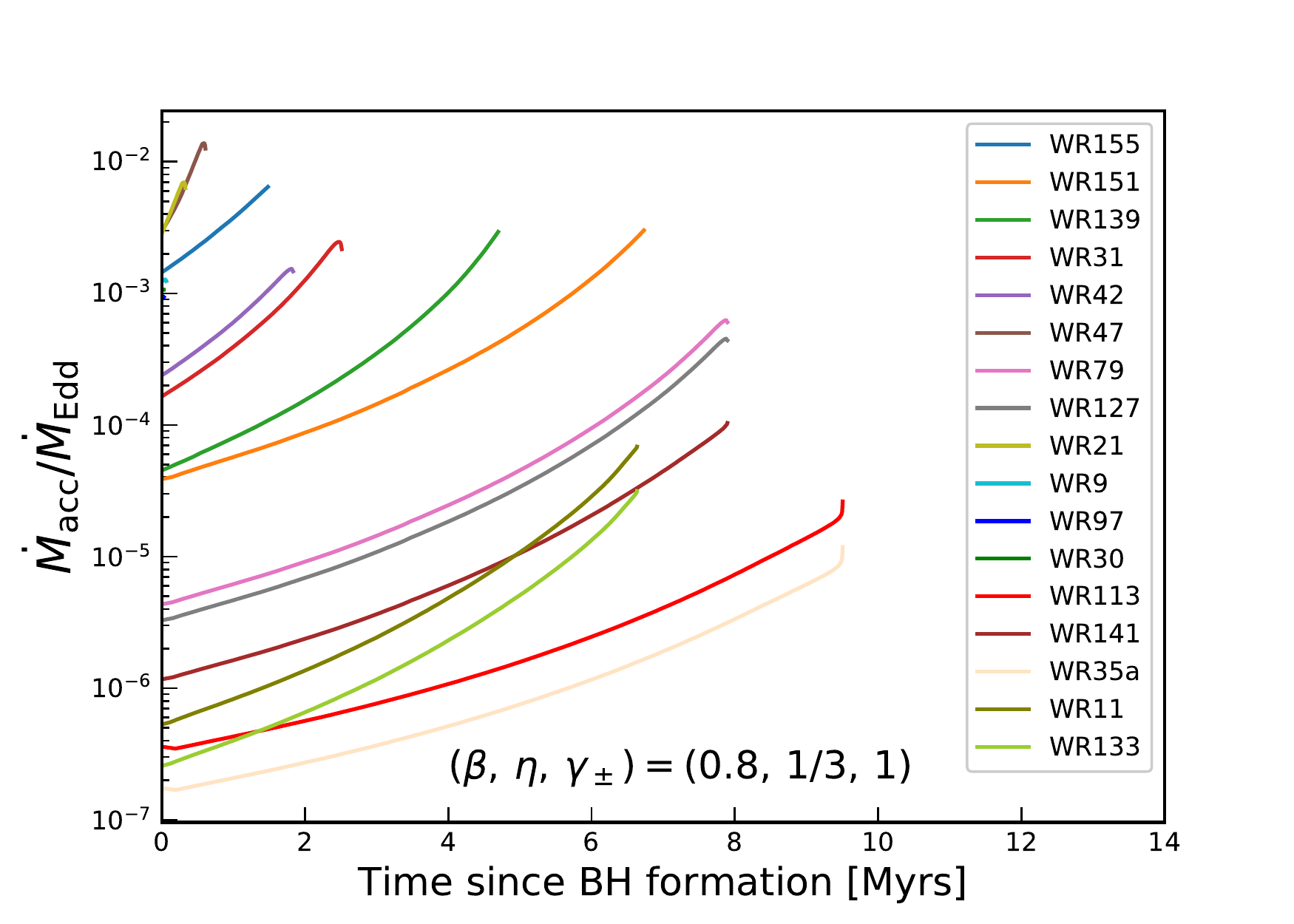}
    \caption{Comparison of mass accretion rate $\dot{M}_{\rm acc}$, mass loss rate of the O star $\dot{M}_{\rm wind}$, and Eddington accretion rate of the BH $\dot{M}_{\rm Edd}$ during the BH+O binary phase modelled from the 17 observed progenitor WR+O binaries, for ($\beta$,$\eta$,$\gamma_{\pm}$) = (0.8,1/3,1). Upper and lower panels presents $\dot{M}_{\rm acc}/\dot{M}_{\rm wind}$ and $\dot{M}_{\rm acc}/\dot{M}_{\rm Edd}$ respectively. The colour coding in the legend denotes the 17 WR+O systems that are expected to form BH+O binaries.}
    \label{mdot_acc}
\end{figure}

Figure \ref{mdot_acc} presents the comparison among the mass accretion rate, 
mass loss rate from the O star, and the Eddington accretion rate of the BH. 
The Eddington mass accretion rate is defined as 
\begin{equation}
    \dot{M}_{\rm Edd}=\frac{L_{\rm Edd}\,R_{\rm ISCO}}{G\,M_{\rm BH}},
\end{equation}
where $\risco$ is the radius of the innermost stable circular orbit around BH,
$G$ is the gravitational constant, $\mbh$ is the mass of BH, 
and $L_{\rm Edd}$ is the Eddington luminosity, evaluated by 
\begin{equation}
        L_{\rm Edd} = L_\odot\frac{65335}{1+X}\frac{\mbh}{\ms},
\end{equation}
where $X$ is the hydrogen abundance in the accreted material,
which is expected to be the hydrogen abundance at the surface of donor star.

The upper panel shows that over 99\% of wind material escape the BH+O system, 
which means the typical timescale of orbital evolution $|a/\dot{a}|$ is longer than
that of mass loss from O star $|\mo/\dot{M}_{\rm O}|$ \citep{IEM2020}.
The mass loss rate of O star is about $10^{-7} - 10^{-6}\,\ms\,{\rm yr}^{-1}$. 
Therefore the orbital period of BH+O binary models can be safely treated as 
constant. The lower panel shows that the $\dot{M}_{\rm acc}$ 
is far below $\dot{M}_{\rm Edd}$. Hence super-Eddington winds from the 
accretor does not occur in our models.

\section{Ratio of O star wind velocity to the orbital velocity}
\begin{figure}
    \centering
    \includegraphics[width=\linewidth]{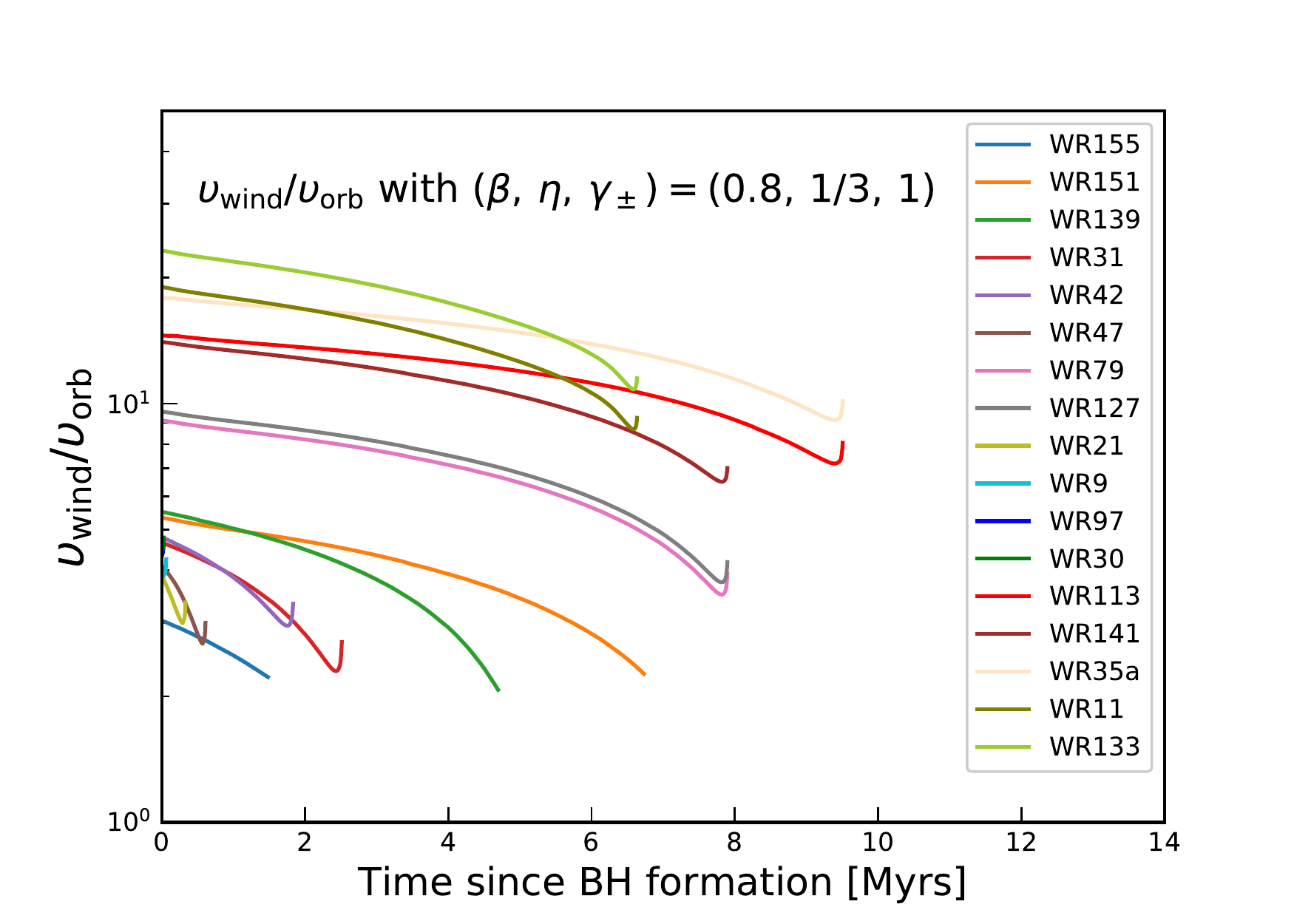}
    \caption{Evolution of the ratio of the O star wind velocity to the orbital velocity during the BH+O binary phase modelled from the 17 observed progenitor WR+O binaries, for ($\beta$,$\eta$,$\gamma_{\pm}$) = (0.8,1/3,1). The colour coding in the legend denotes the 17 WR+O systems that are expected to form BH+O binaries.}
    \label{vw-vorb}
\end{figure}

Figure \ref{vw-vorb} presents the ratio wind velocity divided by orbital velocity $\upsilon_{\rm wind} / \upsilon_{\rm orb}$. The specific angular momentum obtained by SL76 only works in the fast-wind regime ($\vw/\vorb > 1$), which is consistent with our model that we always have $\vw > \vorb$.

\section{Modifications on the disk formation criterion from \citet{iben1996}}

Adopting the specific angular momentum Eq.\,\eqref{specific_j_van} defined by \citet{iben1996} and updated wind velocity Eqs.\,\eqref{vwind} - \eqref{vesc}, the ratio of $\rd/\risco$ is
\begin{equation}
    \frac{\rd}{\risco} = \frac{8}{3}\frac{(\ro/a)^4}{(1+q)^2}\left(\frac{\vorb}{c}\right)^{-2}\left(1+\frac{\upsilon_{\rm wind}^2}{\upsilon_{\rm orb}^2}\right)^{-4}\gamma_\pm^{-1},
    \label{IT96-1}
\end{equation}
where $q$=$M_{\rm O}/M_{\rm BH}$.
Comparing with Eq.\,\eqref{disk-formation-3}, the efficiency parameter $\eta$ for angular momentum accretion is replaced by $(\ro/a)^4$. For the WR+O binaries considered in this work, $(\ro/a)^4$ is much smaller than $\eta$. Hence, we expect that this updated criterion Eq.\,\eqref{IT96-1} predicts fewer wind-fed BH HMXBs than that by Eq.\,\eqref{disk-formation-3}.

Taking the $\beta$ parameter for wind velocity law and the BH spin parameter $\gamma_\pm$ to be equal to 1 in Eq.\,\eqref{IT96-1}, the accretion disk formation criterion (Eq.\,\ref{disk-formation}) can be rewritten as
\begin{equation}
    \frac{R_{\rm O}}{a} \geq \left(2.6\sqrt{1 - \Gamma}\right)^{8/7} \left( \frac{R_{\rm ISCO}}{R_{\rm O}} \right)^{1/7} \left( 1+q \right)^{3/7} \left(1-\frac{R_{\rm O}}{a}\right)^{8/7}.
    \label{correct}
\end{equation}
The wind velocity defined by \citet{iben1996} does not take into account the effect of the Eddington factor on the escape velocity and underestimates the ratio of the terminal velocity to the escape velocity. Further assuming that the binary mass is equal to the mass of the non-compact companion, we obtain 
\begin{equation}
    \frac{\rd}{\risco} = \frac{8}{3}\left(\frac{\ro}{a}\right)^{4}q^{-2}\left(\frac{\vorb^{\prime}}{c}\right)^{-2}\left(\frac{{\upsilon_{\rm wind}^{\prime}}^{2}}{{\upsilon_{\rm orb}^{\prime}}^{2}}\right)^{-4},
    \label{rd_risco_it96}
\end{equation}
where $\vorb^{\prime}$ is the orbital velocity assuming the binary mass is equal to the donor star mass and $\vw^{\prime}$ is the wind velocity defined by \citet{iben1996}, that is Eq.\,\eqref{vw_van}. Combining Eq.\,\eqref{disk-formation} and Eq.\,\eqref{rd_risco_it96} lead to the disk formation criterion obtained by \citet{iben1996} (c.f. eq. 2 of V20),
\begin{equation}
    \frac{R_{\rm O}}{a} \geq  \left( \frac{R_{\rm ISCO}}{R_{\rm O}} \right)^{1/7}  q ^{3/7} \left(1-\frac{R_{\rm O}}{a}\right)^{8/7},
    \label{IT96-2}
\end{equation}
which makes the disk formation much easier than Eq.\,\eqref{correct}. For example, in the BH+O model corresponding to WR 155, with $\Gamma = 0.16$ and $q = 2.5$ at the BH formation time, taking $\gamma_\pm = 1$, $\beta = 0.8$, and the typical O star wind velocity reduces the ratio of $\rd/\risco$ obtained from Eq.\,\eqref{IT96-1} by three orders of magnitudu in comparison to that predicted from Eq.\,\eqref{rd_risco_it96}. 

\end{appendix}

\end{document}